\documentclass[printer]{aa}

\usepackage[varg]{txfonts}
\usepackage{graphicx}
\usepackage{amsmath}
\usepackage{epsfig,amsmath,amssymb}
\usepackage[position=top]{subfig}

\def\ga{\,\hbox{\hbox{$ > $}\kern -0.8em \lower 1.0ex\hbox{$\sim$}}\,}
\def\la{\,\hbox{\hbox{$ < $}\kern -0.8em \lower 1.0ex\hbox{$\sim$}}\,}
\def\beq{\begin{equation}}
\def\eeq{\end{equation}}

\begin{document}

%%%%%%%%%%%%%%%%%%%%%%%%%%%%%%%%%%%%%%%%%%%%%%%%%%%%%%%%
%
\titlerunning{Simulated CMF and IMF in filaments}
\authorrunning{Ntormousi \& Hennebelle}
\title{The core and stellar mass functions in massive collapsing filaments}
\author{Evangelia Ntormousi \inst{1,2} and Patrick Hennebelle \inst{2,3}}
\date{Received -- / Accepted --}

\institute{
Foundation for Research and Technology (FORTH), 
Nikolaou Plastira 100, Vassilika Vouton
GR - 711 10, Heraklion, Crete, Greece \\
\email{entorm@physics.uoc.gr}
\\
\and
Laboratoire AIM, 
Paris-Saclay, CEA/IRFU/SAp - CNRS - Universit\'e Paris Diderot, 91191, 
Gif-sur-Yvette Cedex, France \\
\and
LERMA (UMR CNRS 8112), Ecole Normale Sup\'erieure, 75231 Paris Cedex, France \\
\email{patrick.hennebelle@cea.fr} \\
}

%%%%%%%%%%%%%%%%%%%%%%%%%%%%%%%%%%%%%%%%%%%%%%%%%%%%%%%%

\abstract
{The connection between the pre-stellar core mass function (CMF) and the stellar initial mass function (IMF) lies at the heart of all star formation theories, but it is inherently observationally unreachable.}
{In this paper we aim to elucidate the earliest phases of star formation with a series of high-resolution numerical simulations that include the formation of sinks from high-density clumps. In particular, we focus on the transition from cores to sink particles within a massive molecular filament, and work towards identifying the factors that determine the shape of the CMF and the IMF.}
{We compare the CMF and IMF between magnetized and unmagnetized simulations, and between different resolutions.
In order to study the effect of core stability, we apply different selection criteria according to the virial parameter and the mass-to-flux ratio of the cores.}
{We find that, in all models, selecting cores based on their kinematic virial parameter tends to exclude collapsing objects, because they host high velocity dispersions. Selecting only the thermally unstable magnetized cores, we observe that their mass-to-flux ratio spans almost two orders of magnitude for a given mass. We also see that, when magnetic fields are included, the CMF peaks at higher core mass values with respect to a pure hydrodynamical simulation. Nonetheless, all models produce sink mass functions with a high-mass slope consistent with Salpeter. Finally, we examine the effects of resolution and find that, in these isothermal simulations, even models with very high dynamical range fail to converge in the mass function.}
{Our main conclusion is that, although the resulting CMFs and IMFs have similar slopes in all simulations, the cores have slightly different sizes and kinematical properties when a magnetic field is included, and this affects their gravitational stability.  Nonetheless, a core selection based on the mass-to-flux ratio is not enough to alter the shape of the CMF, if we do not take thermal stability into account. Finally, we conclude that extreme care should be given to resolution issues when studying sink formation with an isothermal equation of state, since with each increase in resolution, fragmentation continues to smaller scales in a self-similar way.}

\keywords{stars:formation }

\maketitle
%%%%%%%%%%%%%%%%%%%----------INTRO--------%%%%%%%%%%%%%%%%%%%%%%%%%%%%%%
\section{Introduction}

Although the process of star formation is by now reasonably understood, the mechanisms responsible for the mass distribution of stars at birth, namely the stellar initial mass function (IMF), have still not been uniquely identified. This is a major challenge for astrophysics, since any variations or environmental dependencies of the the IMF would have significant implications, for instance for the interpretation of galaxy and cluster properties, or for the study of planet formation.

Due to its very definition, the shape of the IMF is actually not easy to reproduce observationally.  Small stars are faint, and therefore hard to resolve in the light of bright, massive stars, so that the low-mass functional behavior of the IMF is not very clear. In addition, even for young stellar clusters close by, crowding of the field is unavoidable. Finally, at the time of the observation, some of the massive stars are already gone, leading to uncertainties in the high-mass end. However, taking these challenges into consideration, the data does seem to agree that, both for stars in the field and in clusters, the stellar IMF peaks at a few tenths of a solar mass, and that its high-mass end follows a power-law distribution: $dN(M)/dlogM\propto \log M^{-1.3}$, where N the number of stars above a mass M \citep{Salpeter_1955,offner2014}. 

Since all known star formation happens in dense molecular cores, the efforts to understand the IMF have shifted towards understanding the origin of the Core Mass Function (CMF) and the fragmentation properties of molecular clouds. In fact, most observational evidence supports the idea that the CMF is shaped like the IMF, with a Salpeter-like power-law distribution at the high-mass end \citep{Testi_1998, Motte_1998, Alves_2007, Nutter_2007,Konyves_2010,offner2014}.  This suggests that, on average, each core forms a star with an efficiency of about 0.1-0.3.  However, given that stars form in multiple systems, and that cores may split or merge during collapse, this idea demands theoretical verification \citep{Holman_2013}.

Indeed, the origin of the CMF and its connection to the IMF have been thoroughly investigated by theoretical work.  \citet{Inutsuka_2001} derived a CMF with a slope close to 2.5 from the gravitational instability of a filamentary cloud, starting from Gaussian perturbations.  \citet{Klessen_2001} instead proposed that, in order to produce a CMF with a slope larger than -2, one needs to take turbulence into consideration.  Along the same lines, \citet{Padoan_2002} obtained a high-mass CMF slope similar to Salpeter by assuming super-Alfv\'{e}nic turbulent fragmentation of the cloud.  \citet{HC_2008} derived the observed stellar IMF from the assumption that the Jeans-unstable cores of a turbulent cloud directly produce the stars. SPH simulations of a collapsing, turbulent molecular cloud by \citet{Smith_2009} led to the conclusion that the pre-stellar core and sink distributions are very loosely connected, and that only the distribution of the first fragments to form in the simulation that resembles the IMF slope (but see \citet{CH10} where the opposite conclusion is reached).  Simulating parsec-sized portions of a molecular cloud formed by supersonic converging flows, \citet{Gong_2015} found that the CMF of all the cores in a single snapshot represented the observed CMF, while high-mass cores were depleted when the cores were selected by gravitational instability.

Recently, the idea that the filamentary morphology of molecular clouds may play a role in the shape of the CMF has emerged following the Herschel Gould Belt survey \citep{Andre_2010, Molinari_2010}.  In this paradigm, clouds fragment into filaments, and filaments into cores.  Even in this context, however, turbulence seems to be an important factor in determining the shape of the CMF \citep{Roy_2015}. \citet{YuehNing2017} integrated the geometry of molecular clouds into the theory of the CMF, proposing that the CMF is a convolution of filament statistics and the individual filament CMF.  In their studies they included the effects of magnetic fields, an element that was often neglected in previous works. 

In this work we look more closely into the role of magnetic fields in the collapse properties of a turbulent, filamentary cloud by means of numerical simulations, also looking into resolution effects.  We study an Orion-sized filament, at isothermal conditions, and focus specifically on the shapes of the CMF and the IMF under different selection criteria. 

We also compare our results with the recent work of \citet{Hennebelle_2017}, who used intense zooming to go from a kpc-scaled box, filled with a stratified interstellar medium, where turbulence was self-consistently driven by supernova explosions, down to a resolution of 400 AU. In the remainder of this paper we will refer to these simulations as FRIGG. This comparison is important because the two sets of simulations are complementary in a number of ways. To begin with, the zooming simulation contains significant core statistics due to the large volume, but does not reach the resolution necessary for the formation of sink particles, which we employ here. Also, although here we set up the collapse of a filament in isolation and not self-consistently, we have control runs without magnetization and can study the effects of introducing a magnetic field. Further, the influence of the 
numerical resolution can be studied since more spatial resolution is being used in this work. Finally, by comparing the two works we can pinpoint the influence of the initial conditions, which are clearly, drastically different from each other.

The plan of the paper is as follows. 
Our numerical method is outlined in Section \ref{numerics}, our results are presented in Section \ref{collapse}, Section \ref{summary} summarizes and discusses our findings, and Section \ref{conclusions} concludes the paper.

\begin{figure*}[h!]
\centering
     \subfloat[Model H]{
      \includegraphics[width=0.45\linewidth]{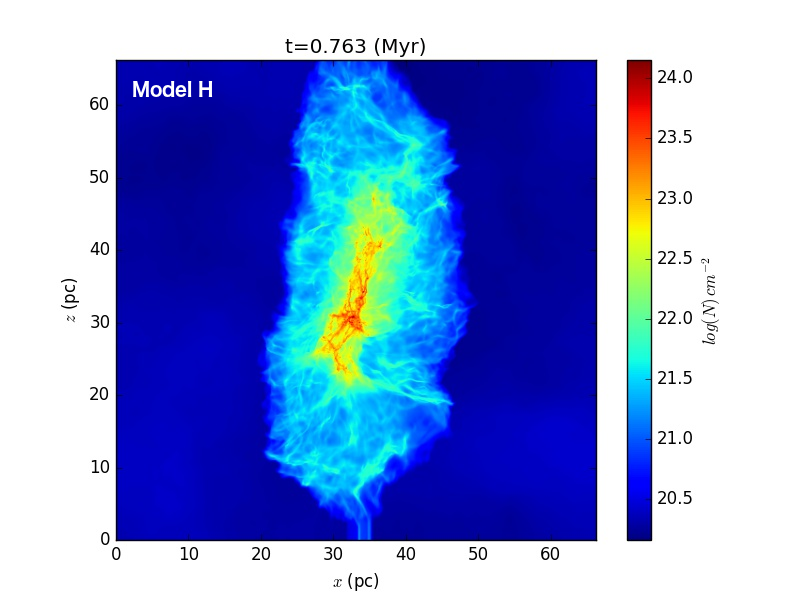}
     }
     \subfloat[Model M]{
      \includegraphics[width=0.45\linewidth]{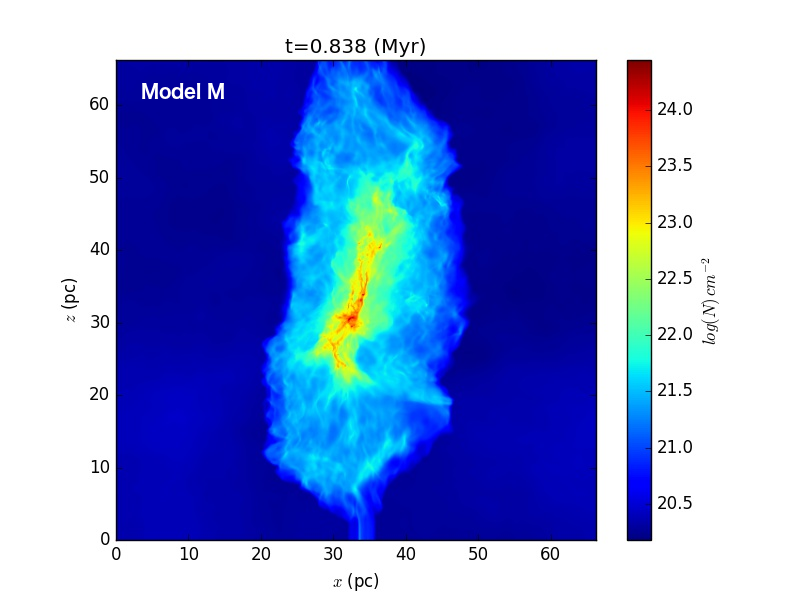}
     } \\
    \includegraphics[width=0.45\linewidth]{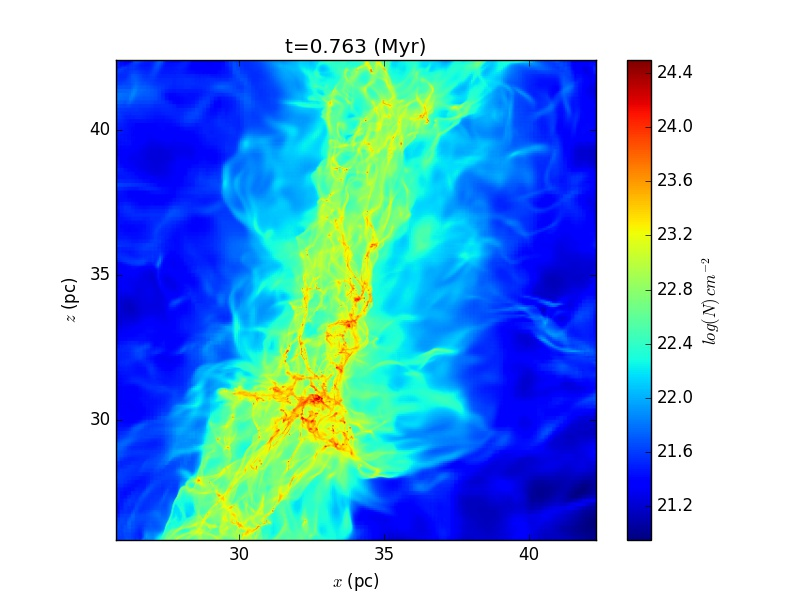}
   \includegraphics[width=0.45\linewidth]{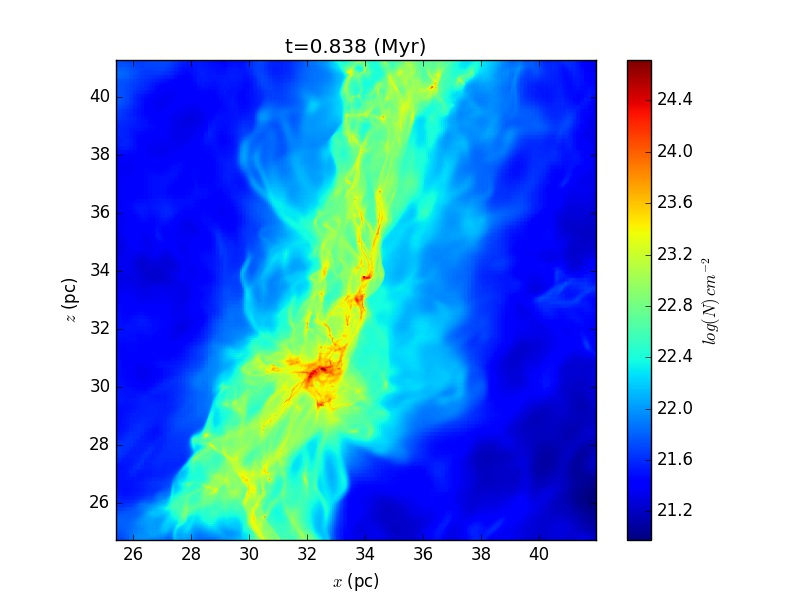}
    \includegraphics[width=0.45\linewidth]{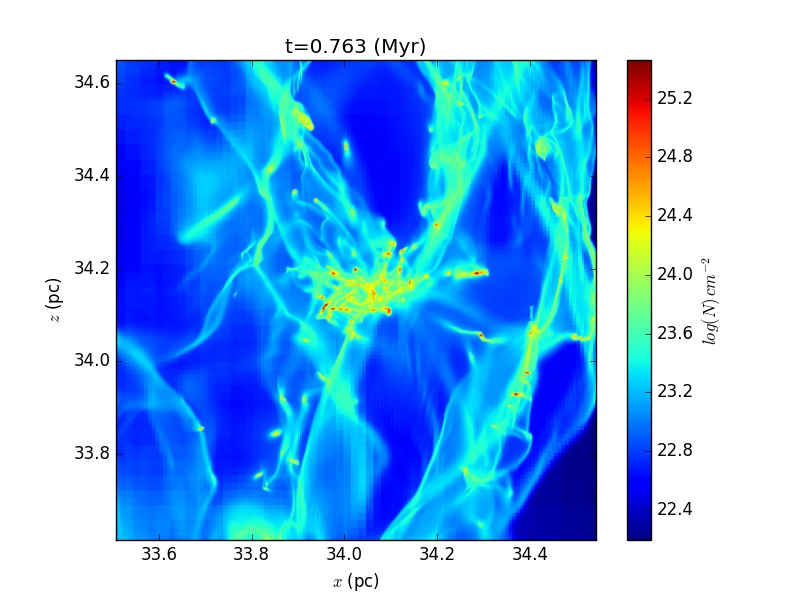}
    \includegraphics[width=0.45\linewidth]{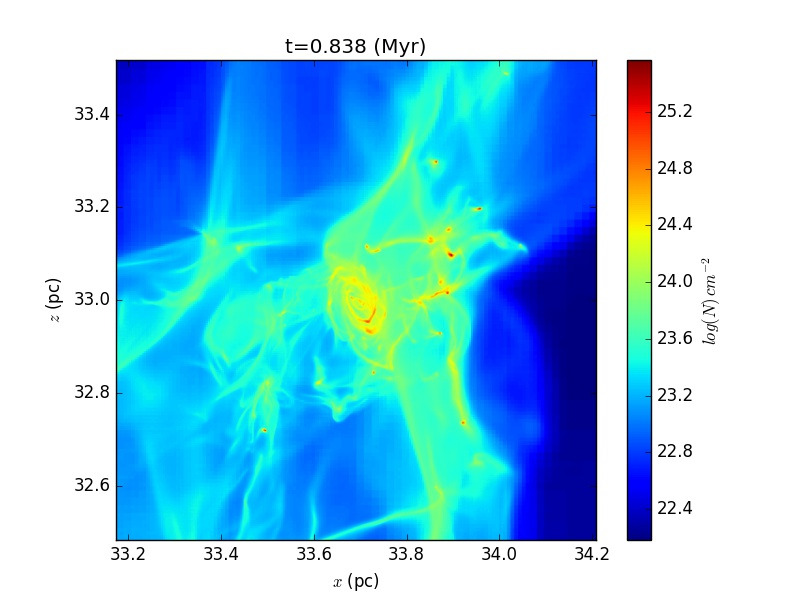}
   \caption{Column density on the xz plane, for Models H and M, gradually zooming into the center (from top to bottom), at reference times HM$_2$.}
     \label{projections_xz}
\end{figure*}

%%%%%%%%%%%%%%%%%%%%%%%%%%%%%%%%%%%%%%%%%%%%%%%%%%%%%%%%%
\section{Isolated filament collapse: Method and setup}
\label{numerics}

\subsection{Numerical code}

We use the publicly available MHD code RAMSES \citep{Teyssier_02, Fromang_2006} to perform 
numerical simulations of a magnetized, turbulent, elongated cloud.  
The RAMSES code solves the MHD equations on a Cartesian grid and has Adaptive Mesh Refinement (AMR) capabilities.

The equations solved by the code are:
\begin{eqnarray}
\frac{\partial\rho}{\partial t} +\nabla(\rho\bf v) = 0 \label{continuity} \\
\frac{\partial\bf{v}}{\partial t} + (\bf{v\cdot\nabla})\cdot\bf{v} + \frac{1}{\rho}\nabla P = -\nabla\phi \label{momentum} \\
\frac{\partial E_{tot}}{\partial t} +\nabla(~(E_{tot}+P_{tot})\bf{v} -(\bf{v}\cdot\bf{B})\cdot\bf{B} = -v\cdot\nabla\phi \label{energy} \\
\frac{\partial\bf{B}}{\partial t} -\nabla\times(\bf{v}\times\bf{B}) = 0 \label{induction} \\
\nabla\cdot\bf{B} = 0 
\end{eqnarray}
where $\rho$ the gas density, ${\bf{v}}$ the velocity, $E_{tot}$ the total energy, $P_{tot}$ the pressure, 
${\bf{B}}$ the magnetic field and $\phi$ the gravitational potential.
The Poisson equation is solved with a Fast Fourier Transform technique at the coarsest level, while a relaxation method 
is used for the finer levels \citep{Teyssier_02}.
%%%%%%%%%%%%%%%%%%%%%%%%%%%%%%%%%%%%%%%%%%%%%%%%%%%%%%%%%%%%%%%%%%%

%
\begin{figure}[h!]
   \centering
    \includegraphics[width=\linewidth]{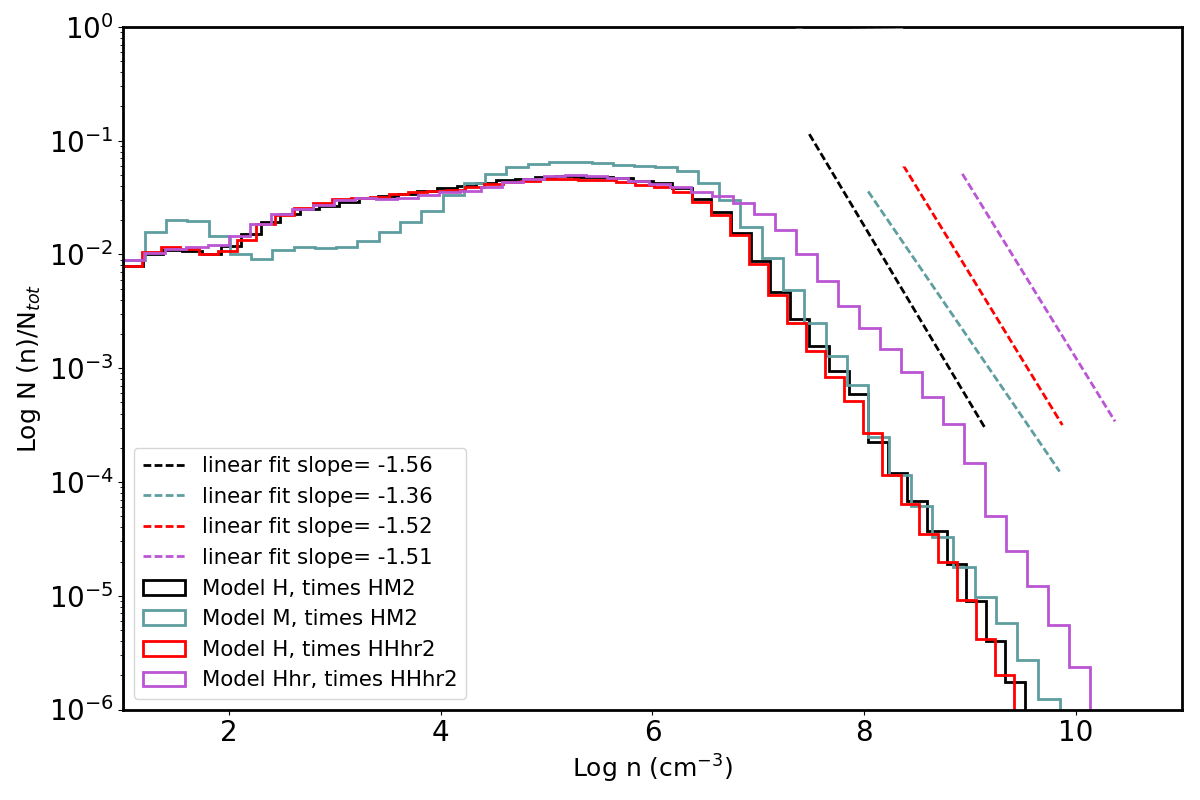}
   \caption{Logarithm of the probability density functions of the density for Models H and M, at reference times HM$_2$, and for Models H and Hhr, at reference times HHhr$_2$. The dashed lines show power-law fits to the high-density gas.}
     \label{densitypdf}
\end{figure}

\subsection{Numerical setup}

The initial setup for all models consists of an elongated ellipsoid of mass $M_{c}=10^{5}~M_{\odot}$, with a radial density profile that follows the relation:
\begin{equation}
\rho(r) = \frac{\rho_c}{1+\frac{(x^2+y^2)}{r_0^2}+\frac{z^2}{z_0^2}}
\end{equation}
where $\rho_c$ is the central density of the cloud, and $(r_0,z_0)$ the radial and vertical size of the inner region of the cloud where the profile flattens. 
The setup accepts a number of parameters which are common among the different models:
\begin{enumerate}
\item The contrast between the density of the cloud edge and that of the ambient medium, set to $C_{cl}=10$, 
\item The initial relation between thermal pressure and gravitation, quantified by $q_{th}=t_{ff}/t_{sound}$, where $t_{ff}$ the free-fall time, and $t_{sound}$ the sound crossing time, set to 0.07, 
\item The initial relation between turbulence and gravitation, quantified by $q_{turb}=t_{ff}/t_{rms}$ where $t_{rms}$ the turbulence crossing time, set to unity.  In this setup the turbulent velocity field has been pre-calculated and read in as an initial condition. It has a Kolmogorov power spectrum, and its amplitude is adjusted to give the desired  $\alpha_{turb}$ parameter. 
\item The ratio between the cloud axes. All clouds have an initial ellipsoidal shape, with two axes of equal length and the third of 2.5 times their length. 
The longest cloud dimension is aligned with the z axis.
\item The equation of state, which here is isothermal with a temperature of about 10~K. 
\end{enumerate}

Based on the above choice of parameters, we calculate the central density of the cloud, $n_0=1500$ cm$^{-3}$, and the length of the major axis,  $L_c$=33~pc. The box size is set to $2L_c$, so that the boundaries are reasonably far from its edges.

These parameters have been chosen to give an initially gravitationally unstable cloud, that nonetheless contains a significant degree of turbulence, much like observed clouds.

We present three simulations: two pure hydrodynamical simulation (Models H and Hhr), and an MHD simulation
where there is an initially uniform magnetic field along the x direction (Model M).    
The ratio of the free-fall time to the  Alfv\'{e}n crossing time of the cloud for model M is $\alpha_{mag}=t_{ff}/t_{A}=0.2$.
Models H and Hhr differ only in numerical resolution, 
and we will use model Hhr for numerical convergence tests.

In these models we use the adaptive mesh refinement (AMR) feature of the code, resolving the
Jeans length always with at least 10 cells.  The coarsest grid is $512^3$ and we activate 7 levels of refinement for models M and H, 
and 8 levels of refinement for model Hhr.  For runs M and H this corresponds to a resolution 
of about $10^{-3}$ pc while for runs Hhr, this corresponds to a resolution of $5 \times 10^{-4}$ pc. 
For reference, this is about 4 and 8 times better than the canonical resolution 
used in FRIGG).  
The properties of the models are summarized in Table \ref{tab:models}.

Finally, star formation is modeled by turning the densest parts of the cloud ($n>10^8$ cm$^{-3}$ in models H and M, $n>10^9$ cm$^{-3}$ in model Hhr) into "sink" particles. A sink is a collisionless component in the code, coupled to the hydrodynamical solver by a Particle Mesh method. 
In this implementation, a gaseous clump will form a sink if it has reached the set density threshold and in addition is virialized, contracting, and does not already contain a sink. The details of the algorithm are described in \citet{Bleuler_2014}. Stellar feedback is not included in these simulations.

Due to the slightly different initial conditions, each of the models takes a different amount of time to reach an initial statistical equilibrium.
For this reason, rather than comparing the models at the same evolution times, we compare them at the time when a similar amount of mass has been accreted onto sink particles. The four comparison pairs are indicated in Table \ref{tab:times}. As a tradeoff between adequate evolution of the dynamics and computational resources, we stop the evolution of the models after approximately one dynamical time. This corresponds roughly to the latest comparison times in Table \ref{tab:times}.

\begin{table}[!ht]
\caption{{\bf{Simulation parameters}}.  The AMR level, l, is defined as the exponent of $2^l$ that determines the number of cells at each side of the box, assuming that the whole box would be processed at that resolution. 
For example, an AMR level $l=7$ means that the whole box would contain $2^7=128^3$ cells if this were a uniform grid simulation.} 
  \begin{tabular*}{\linewidth}{@{\extracolsep{\fill}}lll}
    \hline
    \textbf{Name}  &  \textbf{Magnetic field}  & \textbf{Min and max AMR level} \\ 
    \hline
    \hline
     M  &  5 $\mu$G  & 9-16  \\
     H  &   0  &  9-16 \\
     Hhr & 0  & 9-17 \\
    \hline
  \end{tabular*}
\label{tab:models}
\end{table}
\begin{table}[!ht]
\caption{Reference comparison times for the models.  For the model names refer to Table \ref{tab:models} and to the text.}
  \begin{tabular*}{\linewidth}{@{\extracolsep{\fill}}llll}
    \hline
    \textbf{Reference} & \textbf{Model}  &  \textbf{Time (Myrs)}  & \textbf{Mass in sinks (M$_{\odot}$)} \\ 
    \hline
    \hline
     HM$_1$  &   &  & \\ 
      & M  & 0.63  & 1528  \\
      & H  &  0.55   &  1731 \\
     \hline
     HM$_2$  &   &  & \\ 
      & M  & 0.86  & 8837  \\
      & H  &  0.76   &  8717 \\
     \hline
     HHhr$_1$  &   &  & \\ 
      & H    &  0.59  & 2948 \\
      & Hhr & 1.17 & 2465 \\
     \hline
     HHhr$_2$  &   &  & \\ 
      & H    &  0.69  & 5842 \\
      & Hhr & 1.26 & 5038 \\
    \hline
  \end{tabular*}
\label{tab:times}
\end{table}
% --------------------------- ---------------------------- ----------------------- --------------------------
\section{Filament collapse}
\label{collapse}

Figure~\ref{projections_xz} shows column density plots on the xz plane for models H and M, gradually zooming into the central region of the cloud as indicated by the labels of the x and z axes.
At large scales the cloud structure is similar between the two models. However, there is much more small-scale structure, leading to more fragmentation in model H with respect to model M. A good example is the 0.2 pc roundish structure in the central region of Model M, which is entirely absent in Model H  (bottom panel of Fig.~\ref{projections_xz}). This is a clear indication of the influence of the magnetic field on the distribution of angular momentum.

A frequently used diagnostic of the star-forming state of a molecular cloud is the column density probability density function (pdf). According to theory, the column density pdf of a turbulent cloud should be lognormal, while a gravitationally collapsing gas should show a power-law pdf \citep{Klessen_2000,Kritsuk_2011,Federrath_2013, Tremblin_2014,Lee_2017}.  Indeed, observational studies of Infrared Dark Clouds (IRDCs) do show clear power-law tails in clouds with ongoing star formation \citep{Froebric_2010, Schneider_2015}. 

Here we use the number density pdfs to describe the dynamical state of the cloud, shown in Fig.~\ref{densitypdf} for all the models. The distributions in models H and M are both peaked at roughly $10^5$ cm$^{-3}$, while the increase in resolution causes a small shift in the lognormal peak of model Hhr to higher values.  All models produce a power-law at high densities ($\log (n/cm^{-3}) > 6$), with slopes between -1.3 and -1.6, in agreement with previous numerical studies. For example, \citet{Federrath_2013}, in a parameter study of turbulent environments, found power-laws of slope between -1 and -2 at high densities for turbulence driven by a mix of solenoidal and compressible modes, and a star formation efficiency between 0 and 5 percent.

The presence of the power law here indicates that gravity is acting at high densities, continuously creating new cores and sinks. In fact, as \citet{Klessen_2000} and \citet{Dib_2005} pointed out, these power-law tails should be time-dependent, becoming flatter as more structures become denser and collapse.  

The effect of resolution is visible here, as the power-law in model Hhr extends to higher values than in model H. There is also a noticeable change in the power law at  $\log  (n/cm^{-3}) =8$, where the sink formation threshold is set. The fact that there is gas above that threshold reflects the additional criteria for sink formation and for accretion onto a sink.

%%%%%%%%%%%%%%%%%%%%%%%%%%%%%%%%%%%%%%%%%%%

\subsection{Core properties}

\subsubsection{Identifying cores}

\begin{figure*}[h!]
\centering
      \includegraphics[width=0.95\linewidth]{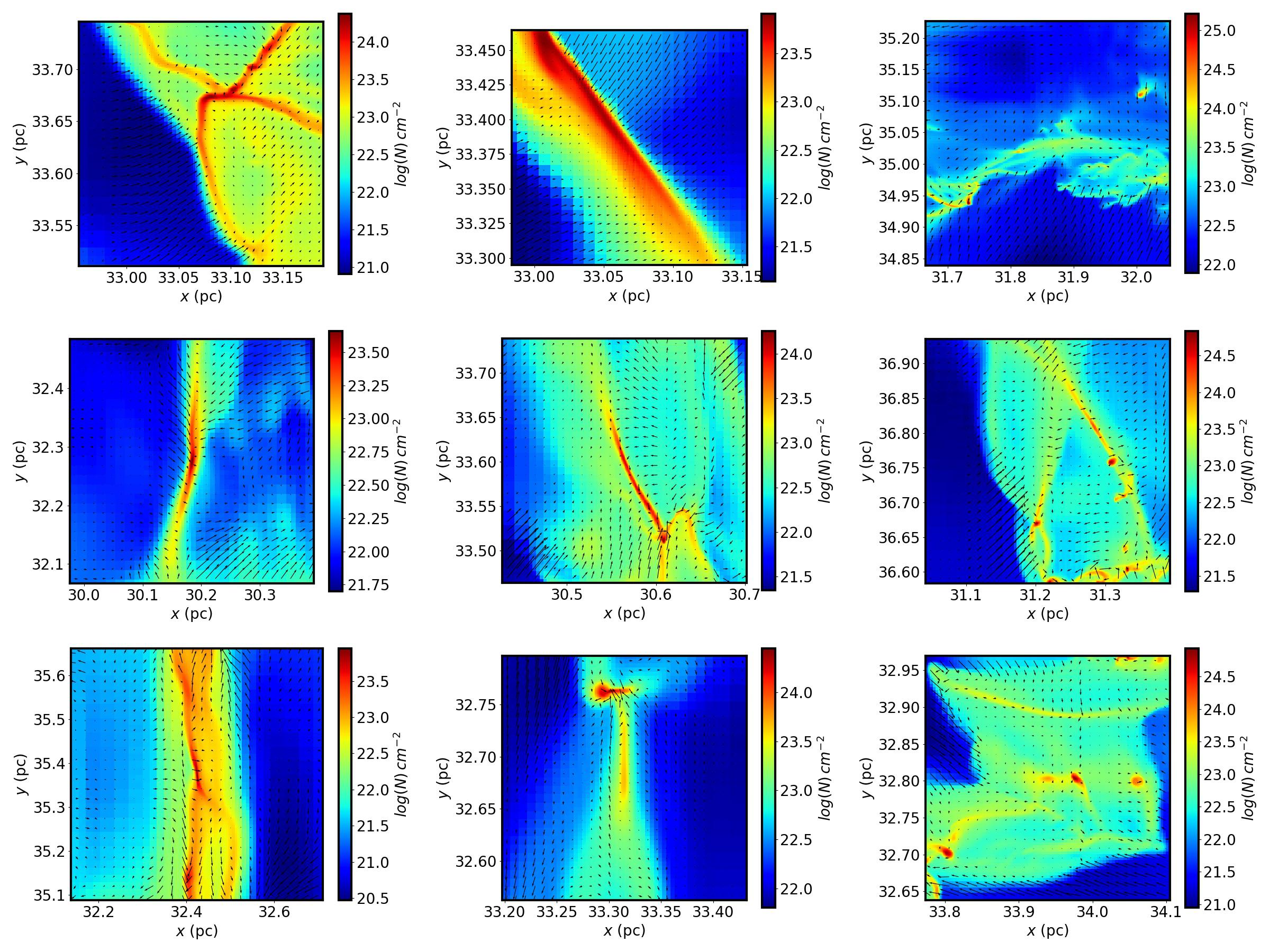}
   \caption{Column density of selected cores on the xy plane, for Model H, at reference time HM$_2$. The black arrows show the projected velocities on the same plane.}
     \label{core_projections_hydro}
\end{figure*}
\begin{figure*}[h!]
\centering
      \includegraphics[width=0.95\linewidth]{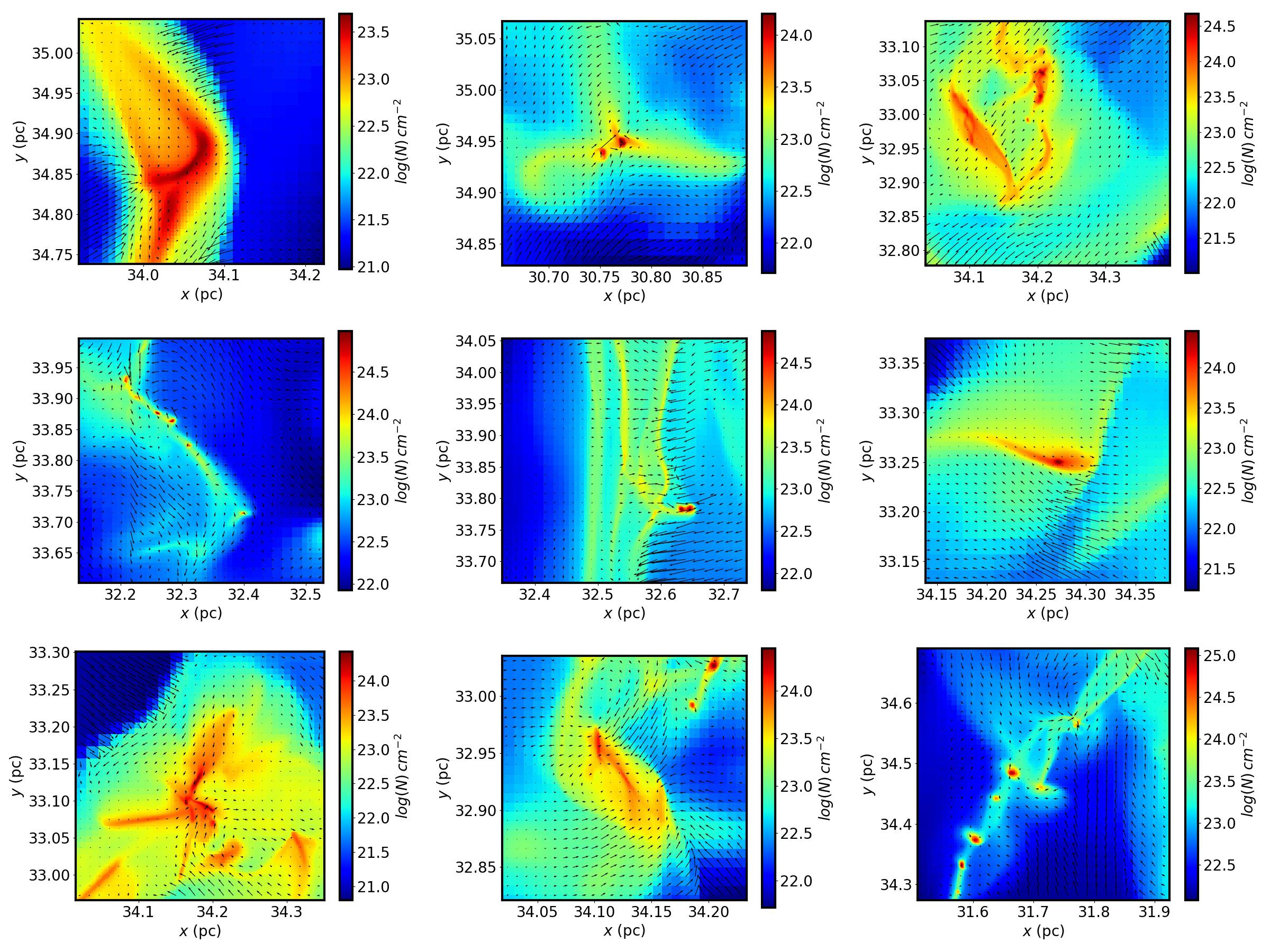}
   \caption{Column density of selected cores on the xy plane, for Model M, at reference time HM$_2$. As in Fig. \ref{core_projections_hydro}, the black arrows show the projected velocities on the same plane.}
     \label{core_projections_mhd}
\end{figure*}
Cores are selected using an implementation of the HOP algorithm \citep{Eisenstein_Hut_98}, which is specifically designed for grouping dense regions.  HOP successively links each grid cell to its densest neighbor, until it reaches a location which is its own densest neighbor.  It then hops to another location.  Eventually, all grid cells that are linked to the same local density maximum are grouped together, excluding locations below a user-defined density threshold (here  3000 cm$^{-3}$).  The algorithm is  only mildly sensitive to the choice of this threshold, since it essentially defines at which contour level the cores will be split. 

Some of the thermally unstable cores identified by HOP in simulations H and M are illustrated in Figs. (\ref{core_projections_hydro}) and (\ref{core_projections_mhd}) respectively, together with their immediate environment, in column density plots on the xy plane. Overplotted on the column density plots is the local velocity field, represented by black arrows. The local velocity is calculated by subtracting the center-of-mass velocity from each map. The axes are numbered in the same way as in Fig.~\ref{projections_xz}. In Appendix \ref{app:core_examples} we show only the locations belonging to the cores, as they are identified by the algorithm.

It is clear that both the density and the velocity structures appear in a variety of morphologies in both simulations. There is a reappearing pattern in both models of cores along filaments, resembling "beads on a string", which is illustrated by two examples at the bottom right panels of  Figs.~\ref{core_projections_hydro} and \ref{core_projections_mhd}. There is also a large number of very elongated cores at sheared regions, examples of which are included in both the aforementioned figures. However, in order to make more general comments on potential differences between models, we have to turn to statistics.

\begin{figure}[h!]
\centering
%
   %\vspace{-0.5cm}
   \subfloat[Model H]{
     \includegraphics[width=0.9\linewidth]{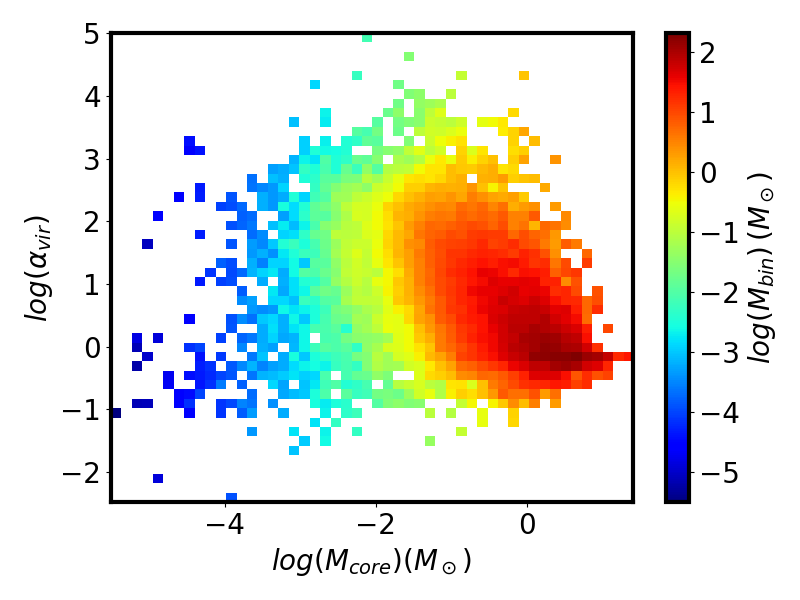}
     }\\
     \subfloat[Model M]{ \vspace{-0.5cm}
       \includegraphics[width=0.9\linewidth]{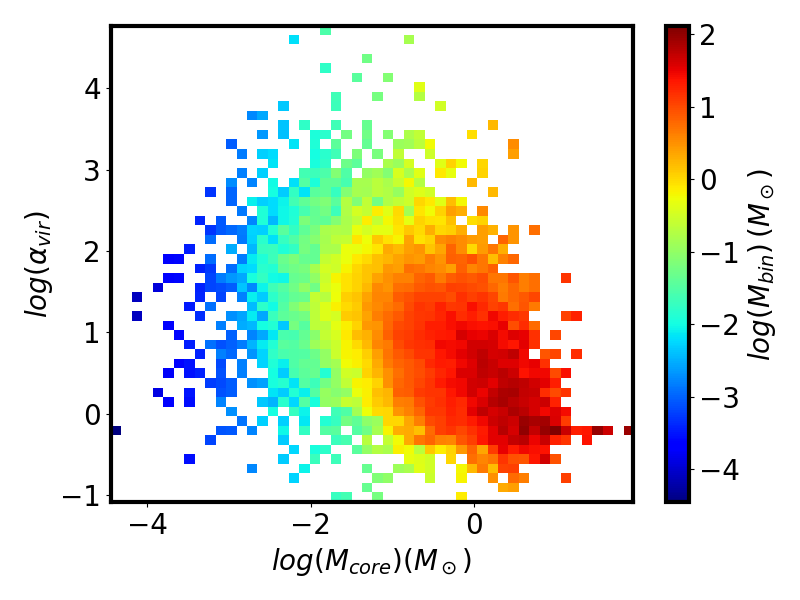}
     } 

   \caption{Virial parameter, $\alpha$ as a function of core mass in Models H  and M, at reference times HM$_2$.}
     \label{core_virial}
\end{figure}
\begin{figure}[h!]
\centering
     \subfloat[Model H]{
     \includegraphics[width=0.9\linewidth]{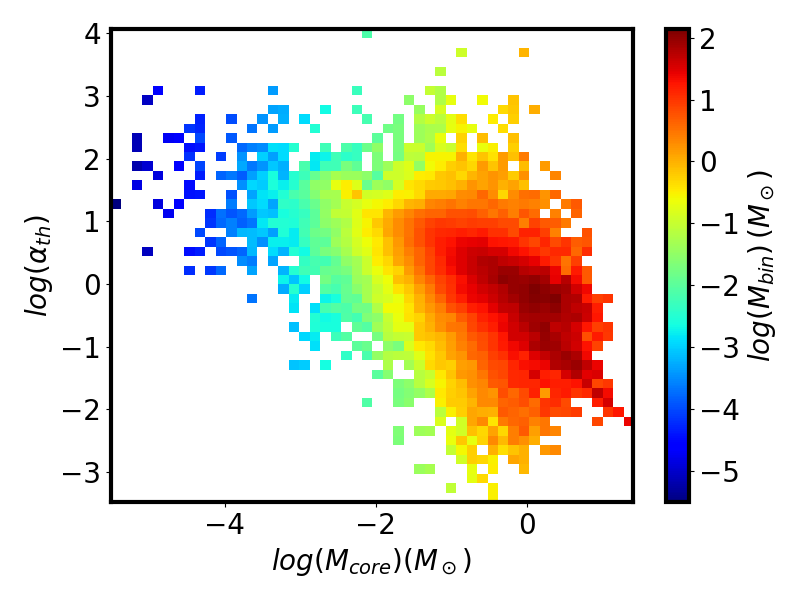}
     }\\
     \subfloat[Model M]{ \vspace{-0.5cm}
     \includegraphics[width=0.9\linewidth]{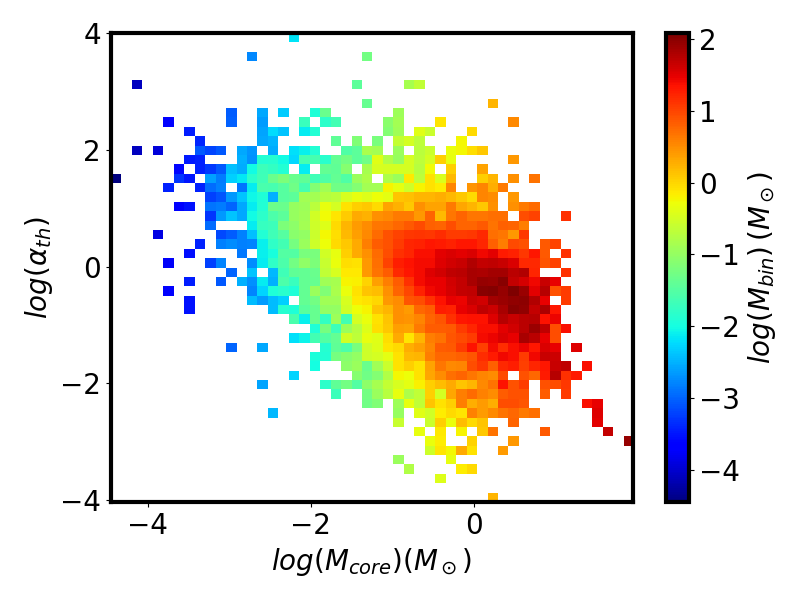}
     } 
   \caption{Mass-weighted 2D histograms showing the thermal virial parameter, $\alpha_{th}$, as a function of core mass in Models H  and M, at reference times HM$_2$.} 
     \label{thermal_virial}
\end{figure}
%
%%%%%%%%%%%%%
\begin{figure*}[h!]
\centering
    \subfloat[Model H]{
     \includegraphics[width=0.45\linewidth]{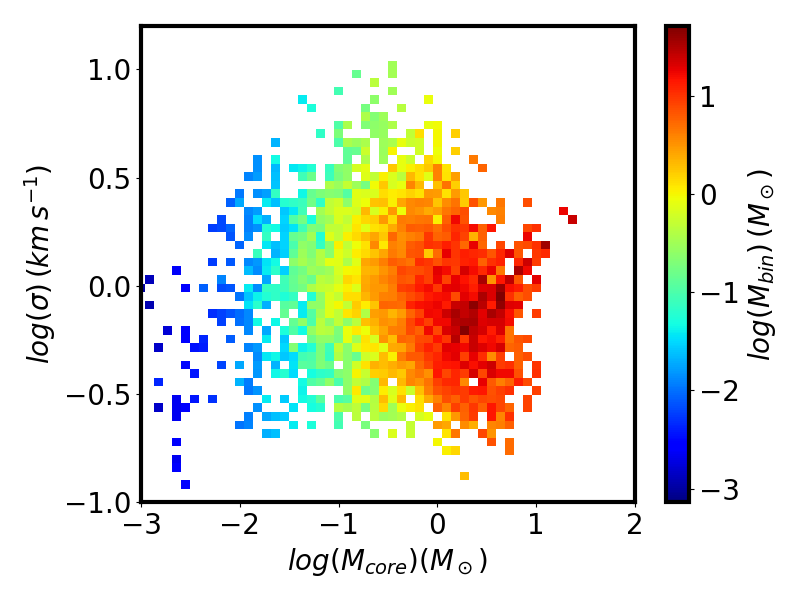}
     }
     \subfloat[Model M]{
       \includegraphics[width=0.45\linewidth]{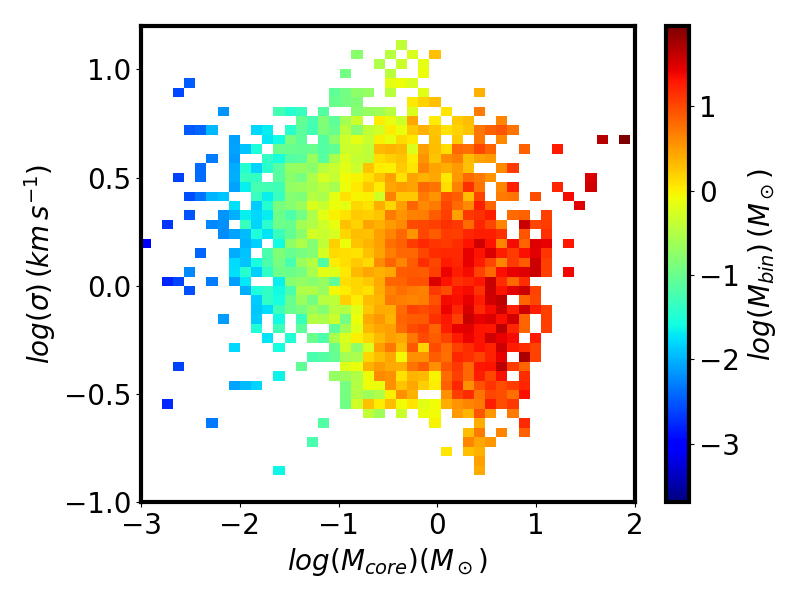}
     } \\
     \includegraphics[width=0.45\linewidth]{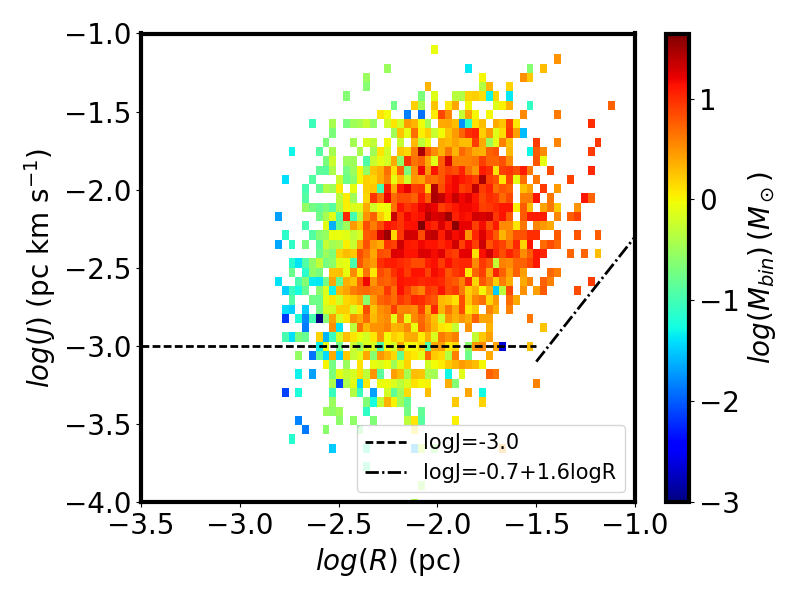}
%     }
       \includegraphics[width=0.45\linewidth]{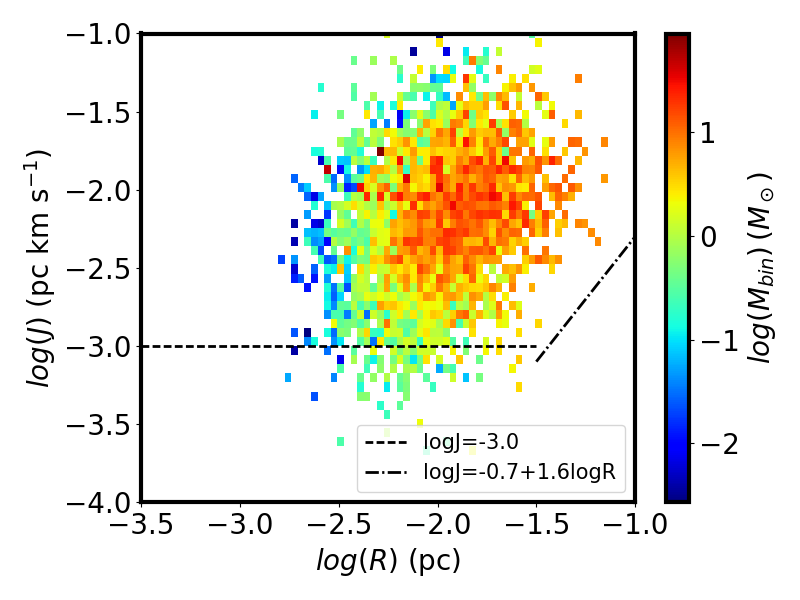}
%     }\\
    \includegraphics[width=0.45\linewidth]{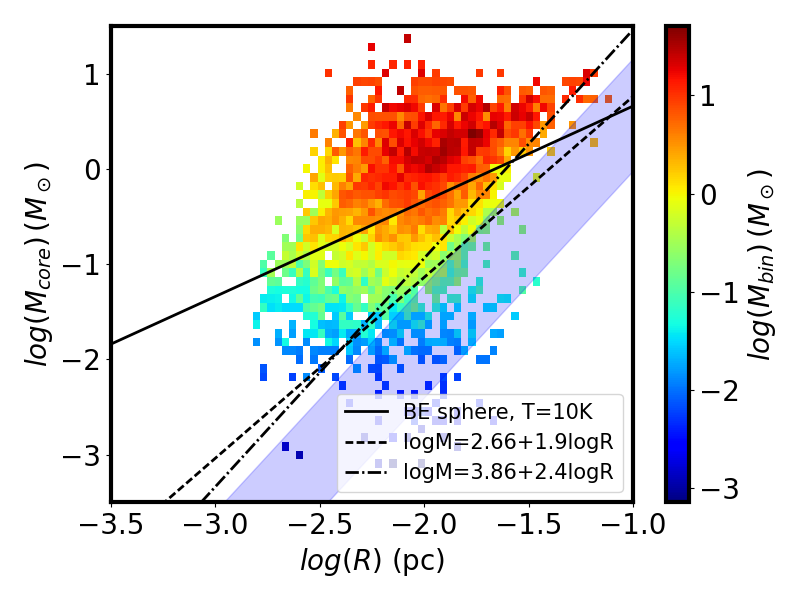}
   % }
      \includegraphics[width=0.45\linewidth]{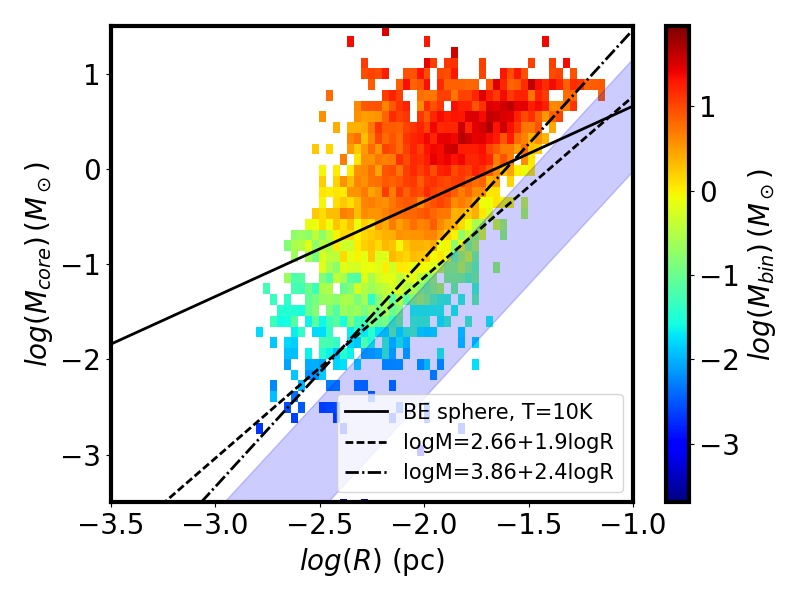}
%    }
\caption{Mass-weighted 2D histograms showing the core properties at reference times HM$_2$. From top to bottom, 3D velocity dispersion as a function of core mass, specific angular momentum as a function of core size, and mass-size relation. Overplotted in the middle panel are the relations between size and angular momentum from \citet{Belloche2002} and \citet{Goodman_1993}, as outlined in \citet{Li_ppvi}. In the bottom panel, the blue shaded band shows the mass-size limits for diffuse CO clouds from \citet{Elmegreen_Falgarone_1996}, the black solid line corresponds to a critical Bonnor-Ebert sphere for $T=10$~K, the dashed line is the \citet{Larson_1981} relation between mass and size, and the dotted-dashed line is the mass-size relation for cores in Taurus quoted by \citet{Kirk_2013} on the \citet{Onishi_1996} data. Only thermally unstable cores are shown.}
     \label{core_prop_1}
\end{figure*}
%%%%%%
%
\begin{figure*}[h!]
\centering
\includegraphics[width=0.49\linewidth]{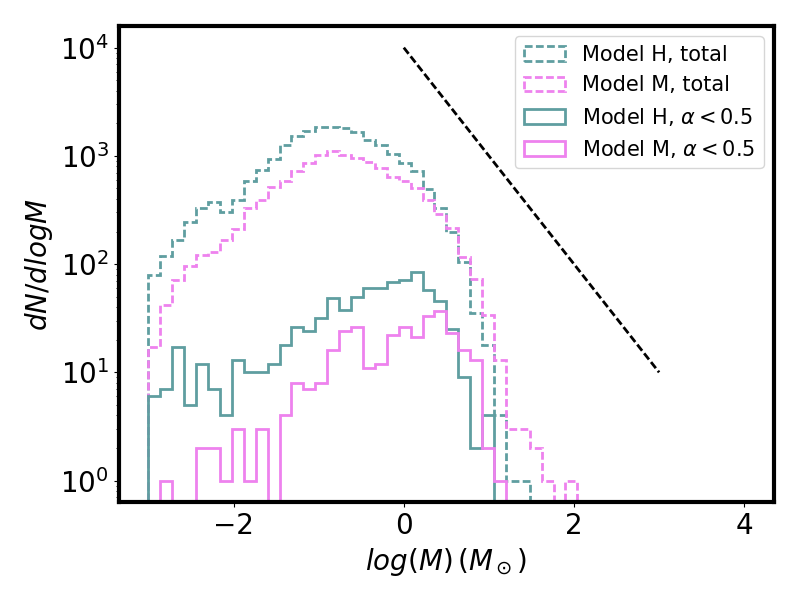}
\includegraphics[width=0.49\linewidth]{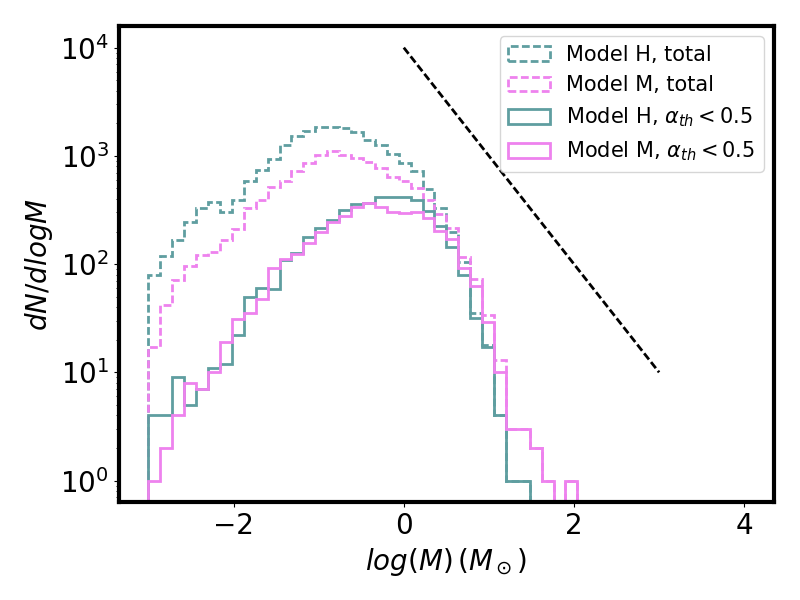}
   \caption{Mass distributions of the cores with $\alpha<$0.5 (left) and $\alpha_{th}<$0.5 (right) for Models H 
 and M, at reference times HM$_2$.  The dashed histograms show the full distribution, while the black dashed line shows the Salpeter IMF slope.}
     \label{cmf_virialcut}
\end{figure*}
%
%%%%%
Since we are interested in studying  cores that will eventually turn into stars, before we calculate the statistical properties for the identified cores, we apply certain selection criteria. For example, we look at the distributions of only the kinetically or thermally unstable cores, as measured by their virial parameter $\alpha = E_{kin}/E_{grav}$ (where $E_{kin}$ and $E_{grav}$ are the kinetic and gravitational energies of the core), and thermal virial parameters, $\alpha_{th} = E_{th}/E_{grav}$ (where $E_{th}$ is the thermal energy of the core).
The thermal and the gravitational energies are direct outputs of the code, so they are readily calculated for each core. The internal velocity dispersion, and from it, the internal kinetic energy of each core, is also fairly simple to calculate after subtracting the center of mass velocity of the structure. 

%%%%%%%%%
\subsubsection{Kinematic and mass-size relations}
The kinetic virial parameters are plotted as a function of core mass in Fig.~\ref{core_virial} for models H and M, in the form of a 2D histogram. Here, the colorbar indicates the total mass in each mass-virial parameter bin.  In both cases, more massive cores have smaller virial parameters, although this tendency is slightly more pronounced in model M.  Moreover, there is a lack of very small-mass (M < $10^{-4}~M_{\odot}$), low-virial parameter cores in model M with respect to model H.

The thermal virial parameters are shown in Fig.~\ref{thermal_virial} as a function of core mass. Since the simulations are isothermal, the thermal pressure support depends on the density, so this relation is much tighter than in Fig.~\ref{core_virial}. On average, the trend is for massive cores to be more unstable to collapse. However, while the typical core in model H has a mass around 1$M_{\odot}$ and a thermal virial parameter around unity, the typical core in model M is slightly larger and more unstable.

It is worth noting that both the kinetic and the thermal virial parameters span a very wide range of values, which partly reflects the wide range of masses and radii. However, a direct comparison of Figs.~\ref{core_virial} and \ref{thermal_virial} shows that in model H there are several very low-mass (log~(M/M$_\odot$) $<-3$ ) cores, which are kinetically unstable ($\alpha<0.5$) but thermally stable. In contrast, model M not only contains on average cores of higher mass, but its low-mass end (log~(M/M$_\odot$)$<-2$) also shows the opposite trend: kinetically stable, but thermally unstable cores.

In Fig.~\ref{core_prop_1} we select only the thermally unstable cores (sub-virial, or $\alpha_{th}<1/2$) and plot their velocity dispersions and specific angular momenta as a function of their mass and radius, respectively, as well as their mass-size relation, in the form of 2D histograms.  Overplotted in the middle panel are the observed size-angular momentum relation from \citet{Belloche2002} (constant angular momentum with radius) and \citet{Goodman_1993} ($j\propto R^{1.6}$), as outlined in \citet{Li_ppvi}. In the bottom panel we show the observed Larson relations \citep{Larson_1981} between mass and size as a dashed line, the mass-size limits for diffuse clouds by \citet{Elmegreen_Falgarone_1996} as a blue shaded band, and the mass-size relation of dense cores in Taurus as quoted by \citet{Kirk_2013} on the \citet{Onishi_1996} data (blue dashed line in fig. 7 of  \citet{Kirk_2013}), as a dotted-dashed line. We have also overplotted as a solid line the critical Bonnor-Ebert mass for a temperature of 10~K, $M_{BE}=2.4~R_{BE}c_s^2/G$, where $G$ the gravitational constant, $c_s$ the isothermal sound speed, and $R_{BE}$ the radius of the sphere, which is also very close to the mean unstable core in the models. Here we define the core size as the largest dimension of the core, similarly to the usual observational definition (defining the radius as the third root of the volume leaves this figure practically unchanged). 

The first thing we notice is that the sizes, masses and specific angular momenta of the cores are very similar between the two runs. Only the three-dimensional velocity dispersions are systematically higher in the magnetized, with respect to the hydrodynamical model. 

In general, the cores in both models have mean kinematic properties consistent with observed trends, but with a much larger scatter.
For example, the velocity dispersion spans almost two orders of magnitude for each given mass, much more than expected for a single cloud.
Such large variations in velocity dispersion are usually reported in observational studies when there are significant variations in the surface density of the clouds studied. In fact, \citet{Heyer_2009}, calculated that, for clouds in virial equilibrium, the velocity dispersion should depend on the column density, leading to a large spread in the velocity dispersion-mass relation. This is also very nicely illustrated in Fig. 1 of \citet{Ballesteros-Paredes_2011}, where the authors plot observed cores from different regions on the same axes. 
Large spreads in velocity dispersions in pre-stellar cores due to a dependence on surface density have also been reported in theoretical work \citep{Ballesteros-Paredes_2002,Camacho_2016}.  
However, the cores in this study are drawn from environments of similar surface density. Therefore, the large scatter in velocity dispersion is propably owed to collapse velocities being interpreted as velocity dispersions.

The dependence of the core angular momenta on their radii that we find here are significantly higher than the relation $j=10^{-0.7}R^{1.6}$ reported by \citet{Goodman_1993}. These authors, however, studied much larger cores and clouds than those we identify here. On the smaller core size, \citet{Ohashi_1999} and \citet{Belloche2002}, suggested there might be a break at around 5000~AU ($\log (R/pc)$ =-1.6), where the relation apparently flattens to $j=10^{-3}~km/sec~pc$, from a constant angular velocity of $\Omega=1.8~km/s/pc$ . Our results seem to be standing somewhere in between: the specific angular momenta scale with radius almost as $R^{1.6}$, but the scatter is so large that they are also consistent with a flat distribution. In both models H and M they are slightly above the constant value $j=10^{-3}~km/sec~pc$ proposed by \citet{Ohashi_1999} and \citet{Belloche2002}, although well within the observed limits of about 10$^{-3}$-10$^{-2}$ km/s pc.

Unsurprisingly, the cores are systematically above the {\citet{Larson_1981}} mass-size relation for diffuse clouds. In contrast, they are in good agreement with, for example, the results on dense cores in Aquila as reported by \citet{Konyves_2015}, and those in Taurus reported by \citet{Kirk_2013} (see also the review by \citet{Hennebelle2012}).

Fron a simulation standpoint, it is interesting that \cite{Chen_2015} find a similar trend ($M_{core}\propto ~R^{2}$) for magnetized cores formed in a post-shock layer, a very different setup than ours.
Finally, these results are in good agreement with what has been inferred in \citet{Hennebelle_2017} for the FRIGG simulations, especially 
the mass-size relation, visible in the top-left panel of their Fig.~3, and the velocity dispersion, portrayed in
the top-right panel of that same figure (see also 
top-left panel of their Fig.~5, as well as their Fig.~6 for the distribution of angular momenta). These are all elatively similar to the results shown in Fig.~\ref{core_prop_1} of the present paper, although with more statistics. In particular, FRIGG contains more massive cores (M > 10 Ms), which are lacking in the present study. On the other hand, here we capture the very small cores, that FRIGG cannot reach due to limited resolution.
The similarities between the two simulations have important implications for both types of models, since FRIGG used very different initial conditions to create the collapsing filaments, namely a stratified, multi-phase ISM, with turbulence created self-consistently from supernova explosions, no sink particles, and a maximum resolution lower by a factor of 4 with respect to this work, triggered by a number of geometrical criteria, plus the local Jeans mass. The similarity of the results indicates that these core properties are not very strongly dependent on the particular choice of initial conditions, as long as the clouds are gravitationally unstable, and that, in large-scale simulations, a resolution of about 400 AU is sufficient to capture the core dynamics. 

\subsubsection{Core mass spectra}

\begin{figure}[h!]
\centering
 \includegraphics[width=\linewidth]{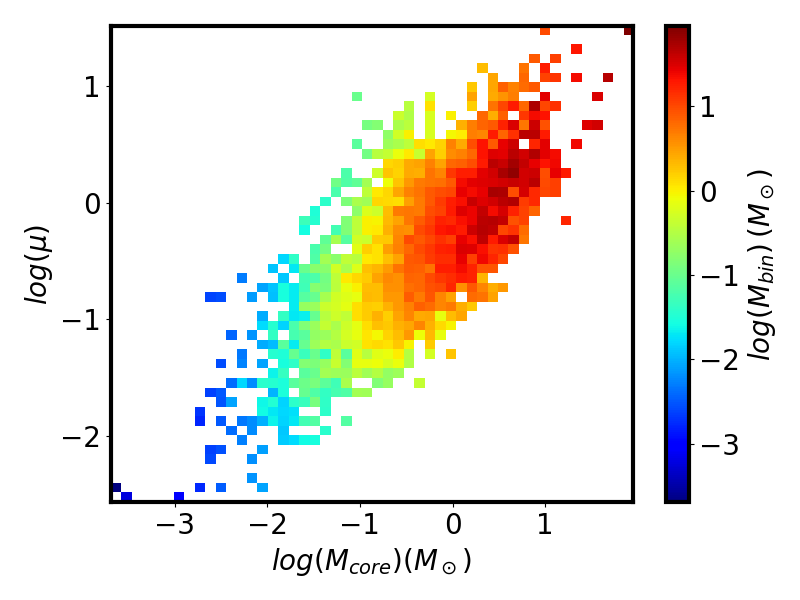}
\includegraphics[width=\linewidth]{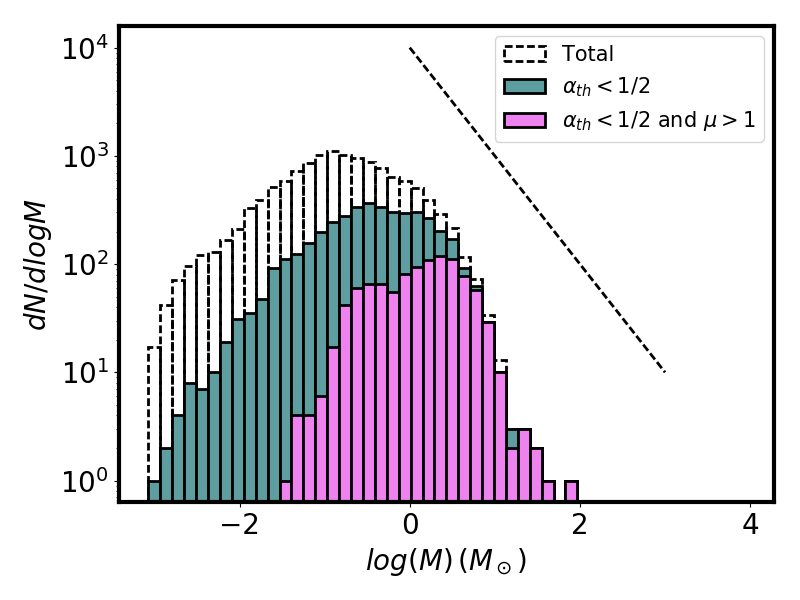}
   \caption{Top: Mass-to-flux ratio, $\mu$, as a function of the mass of the thermally unstable clumps in Model M, at reference time HM$_2$.
Bottom: Mass distribution of the thermally unstable cores with $\mu >$ 1 for Model M at the same reference times, overplotted on the distributions of all cores (black histogram) and cores with $\alpha_{th}<1/2$ (blue histogram), shown in Fig.~\ref{cmf_virialcut}. As in previous plots, the dashed line indicates the Salpeter IMF slope.} 
     \label{mass_mu}
\end{figure}

Given the small differences between runs, it is interesting to compare the total mass distributions of the cores to the mass distributions of only those cores with kinetic, or thermal, virial parameters below one half, namely those more likely to collapse.  The mass histograms of the cores with $\alpha<0.5$ and $\alpha_{th}<0.5$ are plotted in Fig.~\ref{cmf_virialcut}.  In the same plots the histograms of the full mass distributions are drawn with dashed lines.
It appears that selecting cores with a low kinetic virial parameter has a much more dramatic effect on the core mass distribution than selecting them based on thermal stability. This happens because a high velocity dispersion does not only imply internal turbulence, but it can also be due to gravitational contraction. Therefore this criterion possibly excludes also collapsing cores \citep{Traficante_2018}. The effect is more pronounced in run M, where the cutoff affects many high-mass cores.
In contrast, applying a cutoff based on the thermal virial parameter has the effect of moving the peaks of the distributions towards higher values. On close inspection we can see that this shift is larger for model H than for model M, probably because there are still magnetically supported cores among the thermally supercritical distribution (see Fig.~\ref{mass_mu}).

Indeed, we can perform a similar exercise by studying the stability of the cores with respect to their magnetization. This is measured by means of their mass-to-flux ratio, $\mu = M_{core}/\Phi$, where $M_{core}$ the mass of the core and $\Phi$ the magnetic flux in its interior \citep{Mouschovias_Spitzer_1976, Tomisaka_1988}. The mass-to-flux ratios of the thermally unstable cores in model M, normalized over the critical value or collapse according to \citet{Mouschovias_Spitzer_1976}, are plotted in the top panel of Fig.~\ref{mass_mu} as a function of their masses. We can see that the typical thermally unstable core in model M is around a few solar masses and magnetically supercritical. On average, there is almost a linear relation between $\mu$ and core mass. However, there is a significant spread in $\mu$, which varies by almost two orders of magnitude for the same core mass. Again, this result is very reminiscent of the bottom-left panel of Fig.~5 in \citet{Hennebelle_2017}.

The bottom panel of the same figure contains the mass histogram of the thermally unstable cores with a mass-to-flux ratio with values above unity. This histogram, in red, is overplotted on top of the total (black dashed), and the thermally unstable-only (blue) core mass distributions.  
There is a systematic shift of the peak of the distribution towards higher masses as we gradually select cores based on their gravitational instability, confirming the lingering suspicion from Fig.~\ref{cmf_virialcut} that the thermally unstable core selection still contains gravitationally stable cores. In the end, while the total mass distribution peaks at about a tenth of a solar mass, the peak of the magnetically and thermally unstable cores is at a few solar masses. This shows that there is an additional selection for forming stars in model M with respect to model H, which could lead to differences in the stellar initial mass function.

While the general aspects of the core mass spectra are similar to what is presented in 
Fig.~4 of \citet{Hennebelle_2017}, we see that here the peak of the distribution 
occurs at a relatively higher mass: In particular, the thermally unstable core distribution of the present work peaks at about 0.5 M$_\odot$. The corresponding distribution for FRIGG peaks at about 1.5-2 M$_\odot$. This is consistent with the conclusion reached in \citet{Hennebelle_2017} that the peak of the core mass function is determined by the spatial resolution, an issue that we examine further in Section \ref{sec:resol}.

\subsection{Sink properties}
\begin{figure}[h!]
\centering
 \includegraphics[width=\linewidth]{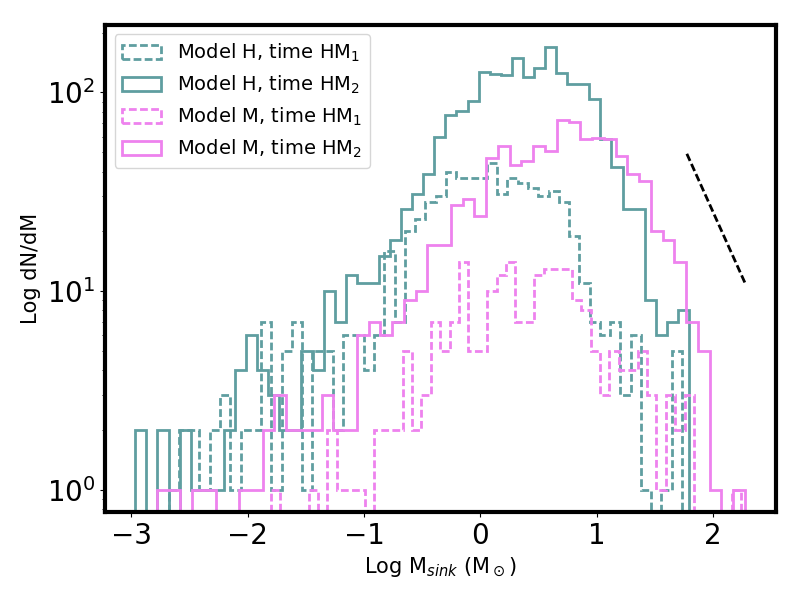}
   \caption{Mass distributions of the sinks for Model H and M, at reference times HM$_1$  and HM$_2$. 
As in previous plots, the dashed line indicates the Salpeter IMF slope.}
     \label{part_histograms1}
\end{figure}

The sink mass distributions for models H and M are plotted in Fig.~\ref{part_histograms1} for both reference times. 
The overall shape of both distributions is similar to that of the CMF, with one clear peak and a power-law slope towards high masses. The slope of the high-mass end in both models is compatible with Salpeter, especially at late times. As with the core masses, model M produces sinks with larger masses than model H, mirroring the relative lack of small-scale structure. However, the peak of the sink distributions of both models is located at higher masses than the respective core mass distributions. 

This puzzling difference probably results from a definition issue: the cores we identify in post-processing are not the same cores that generated the sinks, but those identified at the same time as the sinks. This is similar to observing a cloud where young stars are still located close to their birth sites. 
In fact, based on an analysis not shown in these figures, most of the sinks located close to the clumps are not bound to them, and detach in a timescale of only a few thousand years, below the time resolution of the code output. The few bound sinks are the youngest ones, while the most massive, older ones, that have had time to accrete, have detached dynamically from their parent core.

Let us stress however that the existence of the peak is entirely numerical, and therefore any conclusion regarding its position must be considered with great care. We return to this issue in the following section.

%%%%%%%%%%%%%%%%%%%%%%%%%%%%%%%%%%%%%%%%%%%%%%%%%%%%%%%%%%%
\subsection{Resolution effects}
\label{sec:resol}

\begin{figure*}[h!]
\centering
    \subfloat[Model H]{
     \includegraphics[width=0.45\linewidth]{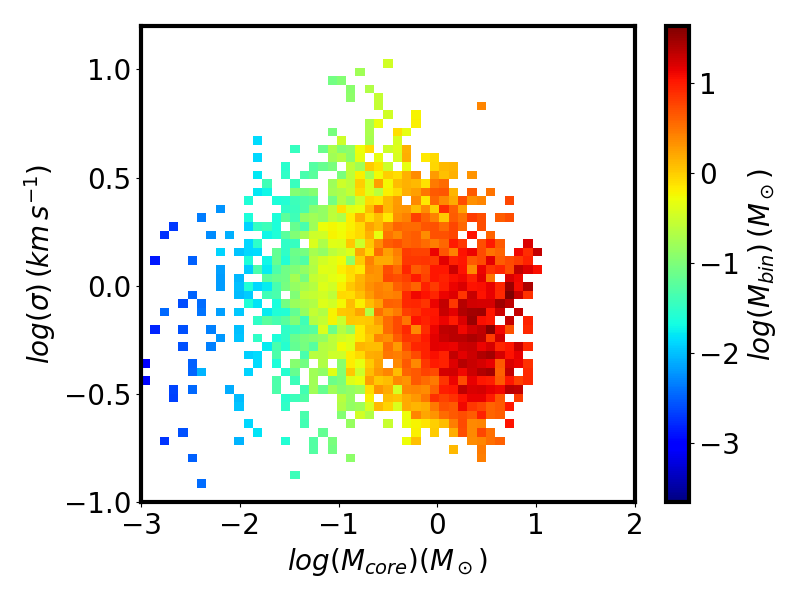}
     }
     \subfloat[Model Hhr]{
       \includegraphics[width=0.45\linewidth]{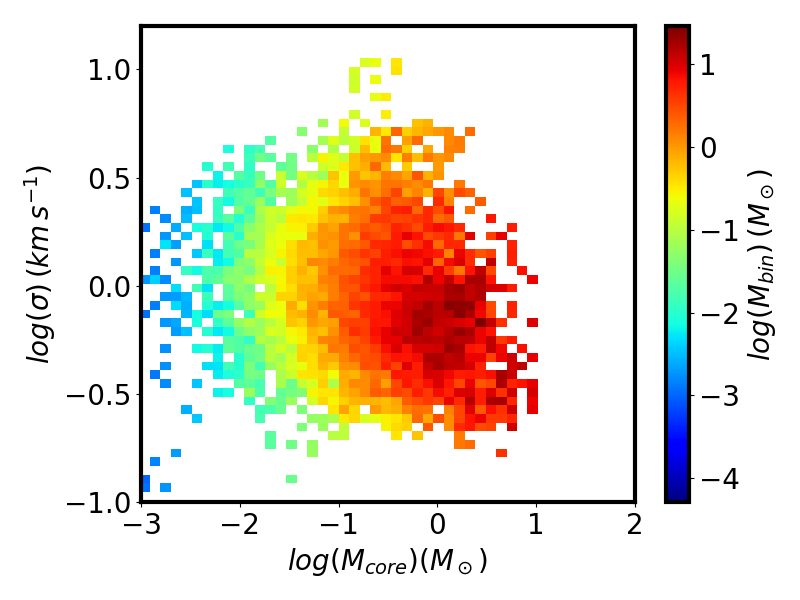}
     } \\
\includegraphics[width=0.45\linewidth]{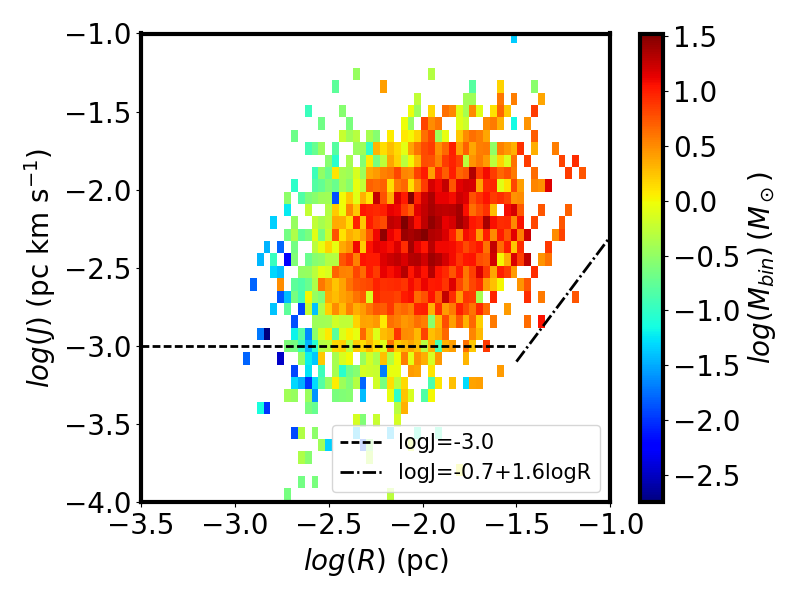}
\includegraphics[width=0.45\linewidth]{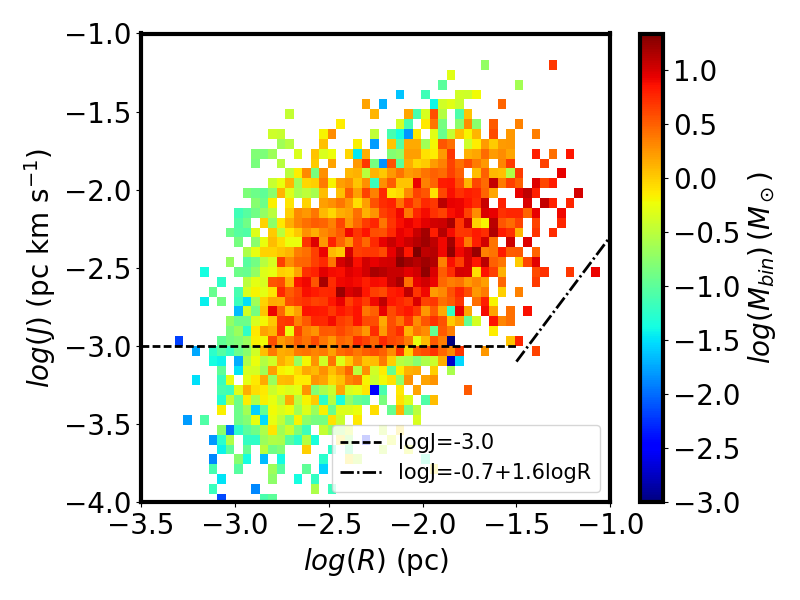}
\includegraphics[width=0.45\linewidth]{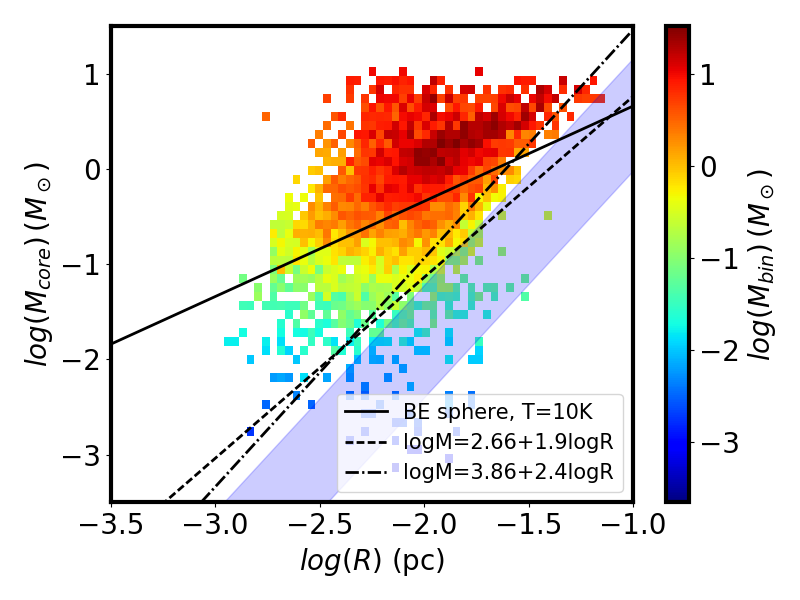}
\includegraphics[width=0.45\linewidth]{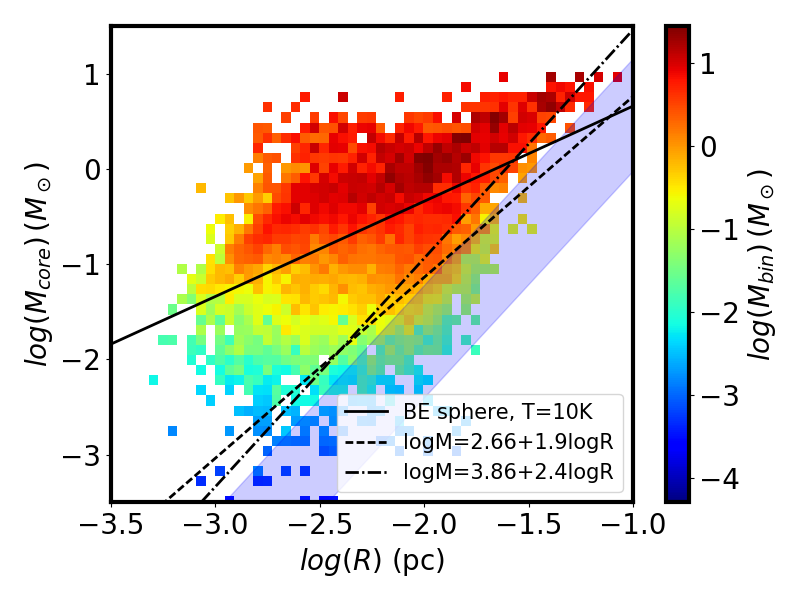}
   \caption{Mass-weighted 2D histograms showing the core properties at reference times HHhr$_2$. As in Fig.~\ref{core_prop_1}, from top to bottom, 3D velocity dispersion as a function of core mass, specific angular momentum as a function of core size, and mass-size relation, with the observed relations as dashed lines, and the mass-size relation for a critical Bonnor-Ebert sphere as a black solid line. Only thermally unstable cores are shown.}
     \label{core_prop_2}
\end{figure*}
\begin{figure*}[h!]
\centering
\includegraphics[width=0.49\linewidth]{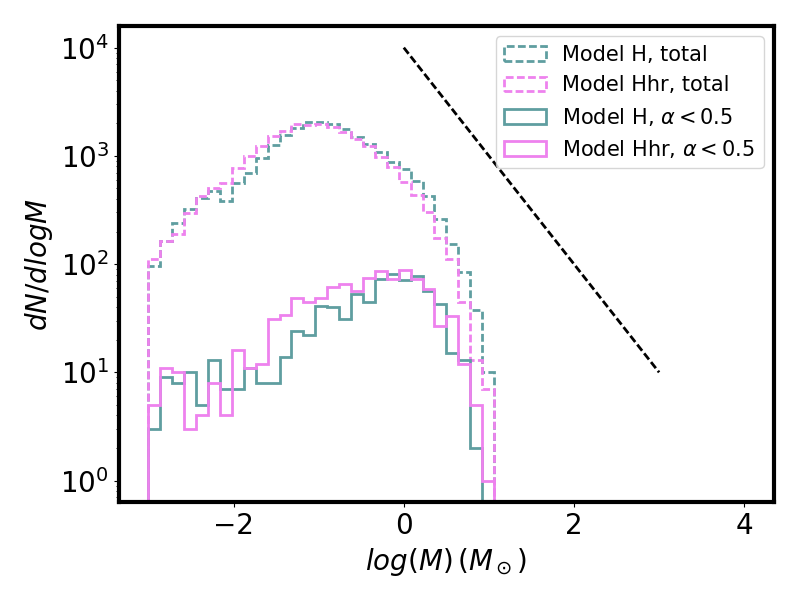}
\includegraphics[width=0.49\linewidth]{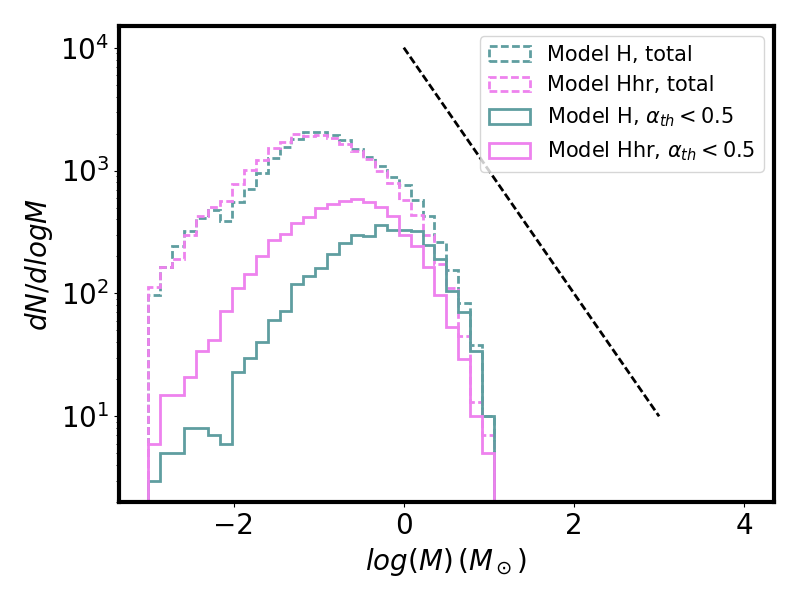}
   \caption{Mass distributions of the cores with $\alpha<$0.5 (left) and $\alpha_{th}<$0.5 (right) for Models H and Hhr, at reference times HHhr$_2$. As in previous figures, the dashed line shows the Salpeter IMF slope.}
     \label{cmf_virialcut2}
\end{figure*}
\begin{figure}[h!]
\centering
\includegraphics[width=\linewidth]{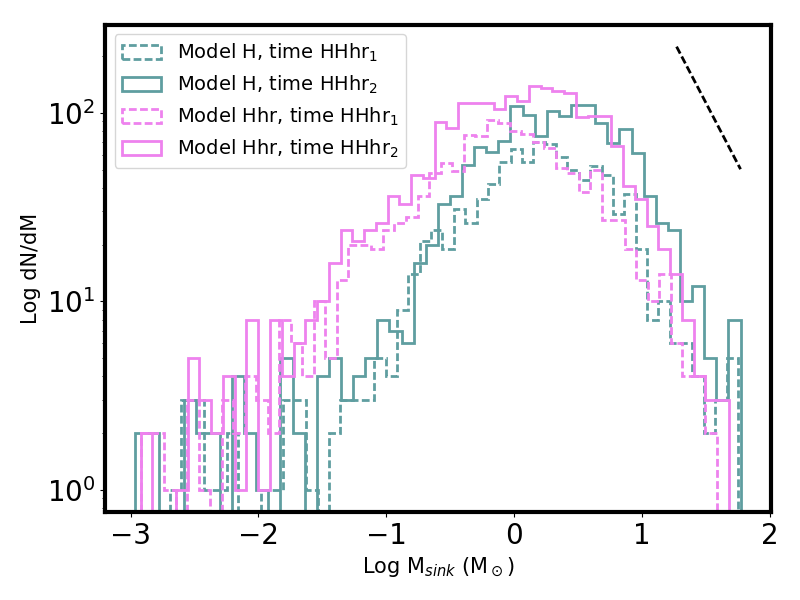}
   \caption{Mass distributions of the sinks for Model H and Hhr, at reference times HHhr$_1$ and HHhr$_2$.}
     \label{part_histograms2}
\end{figure}
The fragmentation of an isothermal cloud under ideal MHD can proceed in ever smaller scales as we increase the resolution.  It is therefore useful to compare the properties of the formed cores and sinks between models H and Hhr and identify any resolution effects.

The core velocity dispersions and specific angular momenta are plotted in Fig.~\ref{core_prop_2}. Overall, there are barely any differences with resolution, which indicates that the self-similarity in mass is reflected in the kinematics of the cores. Increasing the resolution extends all distributions towards smaller values. Interestingly though, the dense cores in model Hhr are closer to the line of critical Bonnor-Ebert spheres. This is probably a result of the fact that their internal structure is better resolved.

Similarly, if we apply a cut at virial and thermal virial parameters as we did previously, we see that the mass histograms of the most unstable cores (Fig.~\ref{cmf_virialcut2}) are very similar between models  H and Hhr, especially the high-mass ends. Only the low-mass populations ($M_{cl}<10^{-1.5}M_{\odot}$) differ strongly between the two models when applying this selection. The reason is that the most massive cores are, in general, the most thermally or kinematically unstable in both cases, and these cores are not affected by the change in resolution. The peak of the core mass function shifts
toward smaller mass although by a factor lower than 2. It is however sufficiently clear to claim that 
numerical convergence is not reached. Higher resolution simulations should be performed to examine whether 
numerical convergence can be achieved regarding the core mass function. 

Finally, resolution also affects the mass distributions of the sinks, apart from better statistics 
(Fig.~\ref{part_histograms2}).  
Both models produce a Salpeter-like slope at late times (bottom panel of the figure), with the peak at slightly lower masses in run Hhr with respect to run H.  
This is in good agreement with the simulations presented in \citet{Lee_2017a} and 
\citet{Lee_2017} where a series of runs with several spatial resolutions has been performed. 
In particular it has been found that with an isothermal equation of state, numerical 
convergence is never reached. The peak of the distribution is determined by the numerical 
resolution and shifts toward smaller mass as the resolution improves. When an adiabatic 
equation of state is used at high density, that is to say when the effective equation 
of state presents an effective polytropic index larger than 4/3, numerical convergence
can be reached. In realistic conditions, this requires a spatial resolution typically smaller
than 20 AU. 

%%%%%%%%%%%%%%%%%%%%%%%%%%%%%%%%%%%%%%%%%%%
\section{Summary and discussion}
\label{summary}

We have presented high-resolution numerical simulations of a massive, collapsing molecular filament, including sink particles, in order to follow the transition from the CMF to the IMF.  We performed one magnetized run and two hydrodynamical runs at different resolutions, in order to study both the effects of magnetization and of resolution on the properties of sinks and cores. 

Our most striking result is that the magnetized model produces unstable clumps and sinks of higher masses than those of the hydrodynamical model. In other words, that the presence of a magnetic field reduces small-scale structure. This finding is in agreement with the work of \citet{Hennebelle_2013}, who showed that in simulations of magnetized, turbulent fluids, magnetic tension helps maintain structures coherent even in the presence of shear stress. Another effect is certainly the extra support that magnetic field provides against gravity. 

In our simulations, the magnetized clumps have higher velocity dispersions than their hydrodynamical counterparts, but similar mass-size relations and specific angular momenta.  This is a clear indication that the magnetic field alters the way momentum is channeled from large to small scales in a non-linear way, but the common ingredients among these models, namely gravity and turbulence, play a dominant role in the statistical kinematic behavior of the cores.  
However, as \citet{Seifried_2015} pointed out, the initial relative orientation of the magnetic field with respect to the filament main axis is a parameter that can alter the fragmentation properties of the filament altogether, and is one that we have not explored in this work.

In order to identify the processes that shape the IMF, we examined the effect of certain selection criteria on the mass distribution of the cores.  
A selection by virial parameter ($\alpha<0.5$) is the most impactful, in the sense that it removes a lot of cores from the total distribution. However, such a selection removes also the collapsing cores, since they host high internal velocities.  Therefore, although the way kinetic energy is imparted to the cores is crucial in determining their stability, and therefore also the IMF, it is hard to determine the virial state of the cores by measuring the velocity dispersion.  

Of course, some effects have been neglected in the calculation of the virial parameter that could potentially alter the results.  For example, \citet{Kirk_2017}, based on their findings on Orion A, suggest that the pressure of the ambient cloud has to be taken into account when calculating the virial state of the cores, something we have not included in this analysis.

Selecting gravitationally unstable cores by means of their thermal virial parameter or ($\alpha_{th}<0.5$) is more successful in excluding the low-mass cores. This causes the peak of the distribution to move towards higher mass values. 
However, magnetic support needs to be taken into account as well. Interestingly, on average the logarithm of $\mu$ scales broadly linearly with mass. In general, the mass-to-flux ratio of the thermally unstable, magnetized cores, spans a wide range of values, including many subcritical cores. Therefore, a further selection according to magnetic criticality is necessary in order to isolate the collapsing cores. 

In all runs, the sink mass function of both models resembles the observed IMF, but peaks at higher masses, which is clearly a resolution effect. However, at the same resolution, model M produces sinks with typically a factor of two higher masses than model H. One possible reason for this that magnetization stops core fragmentation, either by reducing the small-scale structure, either by imparting larger amounts of angular momentum to the cores. 

The typical core mass in our models is lower than the typical sink mass, independently of magnetization or resolution. In fact, only a few sinks are still bound to the clumps at any given time, and are those with ages typically below a few thousand years. 
This is somewhat surprising, because it indicates that sinks can accrete, grow and detach from their parent core very fast.  On the other hand, stellar feedback processes are not included in this work, and could change this finding drastically.

We can gain some intuition into the matter of sink detachment by looking at the work of \citet{Gong_2015}, who studied the formation of cores and sinks in converging flow environments, focusing on small regions (1~pc) and therefore achieving higher resolution than our simulations. In their study they distinguish between the instantaneous core distribution, which they find to be close to the observed, and the mass distribution of the cores at their individual collapse time, which shows a deficit of massive cores, concluding that massive cores continue to accrete after their collapse time. Although we do not make this distinction in core selection, this early detachment of the sinks from the cores means that cores continue to exist and accrete after they have already formed stars. 

Our simulations fall in the range of scales between galactic simulations of the ISM and small-scale, core collapse simulations, so it is instructive to compare to works covering larger and smaller scales. Throughout the text we have been comparing in detail to the results of the FRIGG simulations from \citet{Hennebelle_2017}, who present trends similar to this work in terms of the virial parameters, velocity dispersions, mass-to-flux-ratios and angular momenta of the cores.
Another state-of-the-art project involving zooming simulations is the SILCC project \citep{Walch_2015}, where a large portion of a galaxy is simulated with different supernova feedback recipes, chemistry, and magnetic fields. In particular, in \citet{Girichidis_2018} the authors discuss the role of magnetic fields in the evolution of molecular clouds, in simulations with a maximum resolution of 1~pc. Their Fig.~23 shows the mean mass-to-flux ratio of the clouds in different simulations as a function of time, which at 60~Myrs of evolution agrees very well with the distribution we obtain for a single cloud (Fig. \ref{mass_mu}).
Since our initial filament is set up as gravitationally unstable, and the finest mesh refinement in these large-scale simulations is triggered by the local Jeans instability, these similarities with large-scale simulations imply that the core properties do not depend strongly on the specific mechanism that drives the turbulence on large scales, or on the presence of sinks, but probably on the gravitational instability of the clouds.

On the other hand though, \citet{Kuffmeier_2017} performed zoom-in simulations from a full molecular cloud scale down to sub-AU sizes and found a variety of cores, the star-forming properties of which depend on the magnetic and velocity field of their environment. This is an indication that at sub-AU scales there could be other processes at work that homogenize or differentiate the products of star formation, and are not taken into account in any of these works, such as radiative feedback, and magnetic diffusion.

Again on the small-scale end, \citet{Lee_2017} studied the collapse of small clouds with a polytropic equation of state, which allowed them to achieve convergence in the IMF. They found that cold cloud collapse simulations with an isothermal equation of state the peak of the IMF constantly shifts to smaller values, which confirms our observations.   

\section{Conclusions}
\label{conclusions}
Our main conclusion is that a magnetic field affects the gravitational collapse behavior of a turbulent massive filament in a complex way. In particular, we find that:

\begin{enumerate}
\item A magnetized filament is less fragmented, and hosts cores with slightly higher internal velocity dispersions than a hydrodynamical filament.
\item Otherwise, the cores in the two models are very similar. Their specific angular momenta and their mass-size relation are similar, and all in line with the observed values.
\item The CMF of a magnetized filament, containing only thermally and magnetically supercritical cores, peaks at higher values than the CMF of a hydrodynamical filament, containing only thermally supercritical cores.
\item In the magnetized run, the thermally unstable cores have a wide distribution on mass-to-flux ratios $\mu$, but on average, the logarithm of their masses relates almost linearly to the logarithm of their $\mu$.
\item The sink mass distribution of a magnetized filament peaks at higher values than the corresponding distribution of an unmagnetized filament.
\item Sink particles detach dynamically from the cores that formed them only a few thousand years after their formation.
\end{enumerate}
These results have important implications for any star formation theory, because they indicate that magnetization cannot be ignored in the understanding of the CMF and IMF, especially with respect to the core stability.

In terms of numerical convergence, we find that the velocity dispersions and specific angular momenta of the cores are consistent between the two hydrodynamical simulations at different resolutions. They are also similar to those found in large-scale MHD simulations.  However, increasing the resolution does shift the peak of the CMF and of the sink mass function to lower values. We conclude that in high-resolution collapse simulations it is essential to include the relevant small-scale physics, such as stellar feedback, or a different equation of state at higher densities in order to achieve convergence.

Of course, there is a number of factors that could affect these conclusions and merit further investigation. As an example, stellar feedback could stop sink accretion very rapidly, and interact with the magnetic field of the filament. In addition, a further investigation into the topology and strength of the magnetic field is needed in order to draw more general conclusions.
%%%%%%%%%%%%%%%%%%%%%%%%%%%%%%%%%%%%%%%%%%%

\emph{Acknowledgments}

We would like to thank the anonymous referee for their very useful recommendations.
This research has received funding from the European Research Council under the European Community's Seventh Framework Programme (ERC Grant "MAGMIST" FP7/2007-2013 Grant Agreement no. 306483 and ERC Grant Agreement "ORISTARS", no. 291294). 

\bibliographystyle{apj}
\bibliography{orion}
\label{lastpage}

%%%%%%%%%%%%----APPENDIX----%%%%%%%%%%

\appendix

\section{Core definition}
\label{app:core_examples}

\begin{figure*}[h!]
\centering
      \includegraphics[width=0.95\linewidth]{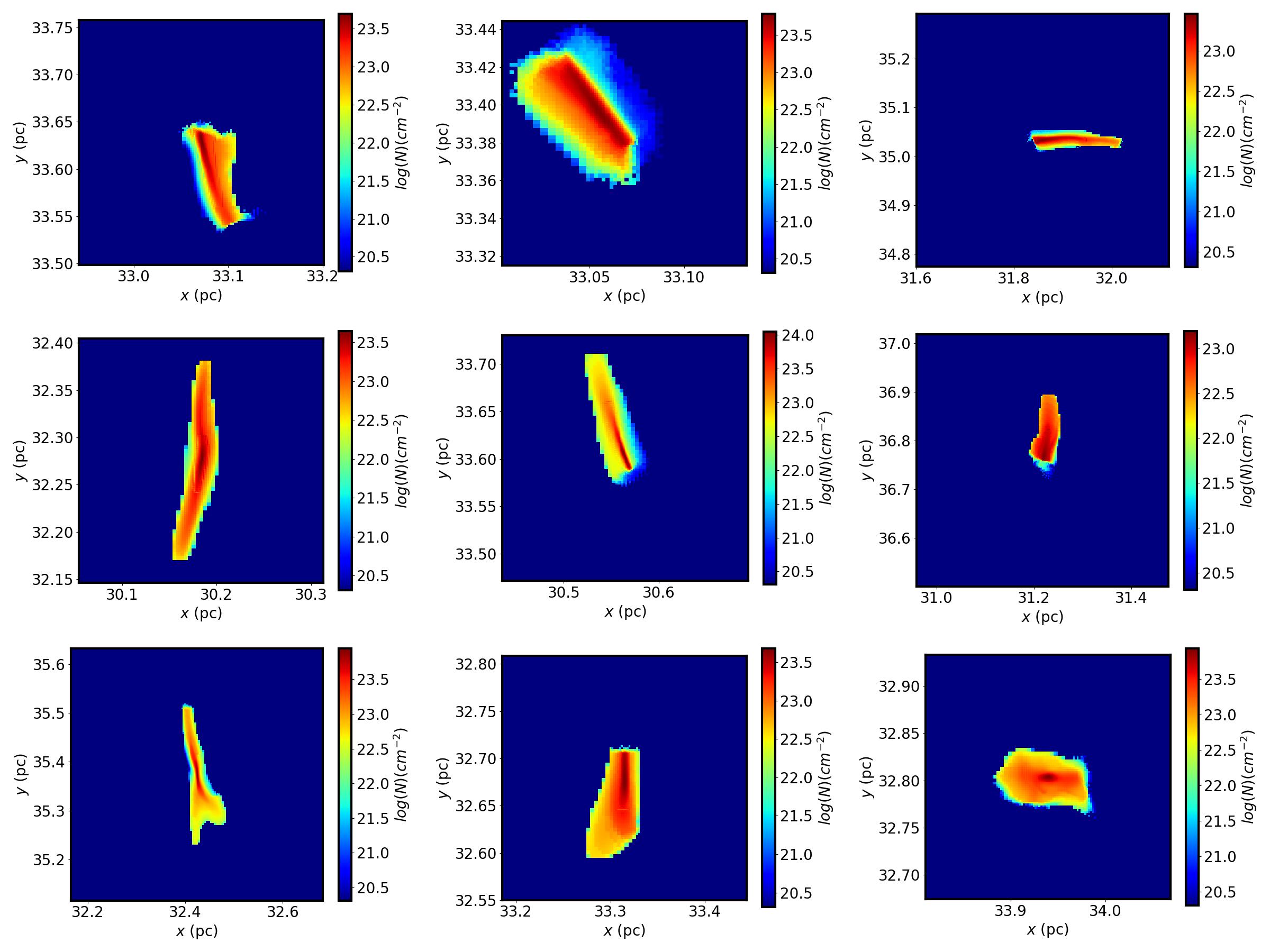}
   \caption{Column density of the cores shown in Fig. \ref{core_projections_hydro}, showing only the locations selected by the HOP algorithm.}
     \label{hierar_hydro}
\end{figure*}

\begin{figure*}[h!]
\centering
      \includegraphics[width=0.95\linewidth]{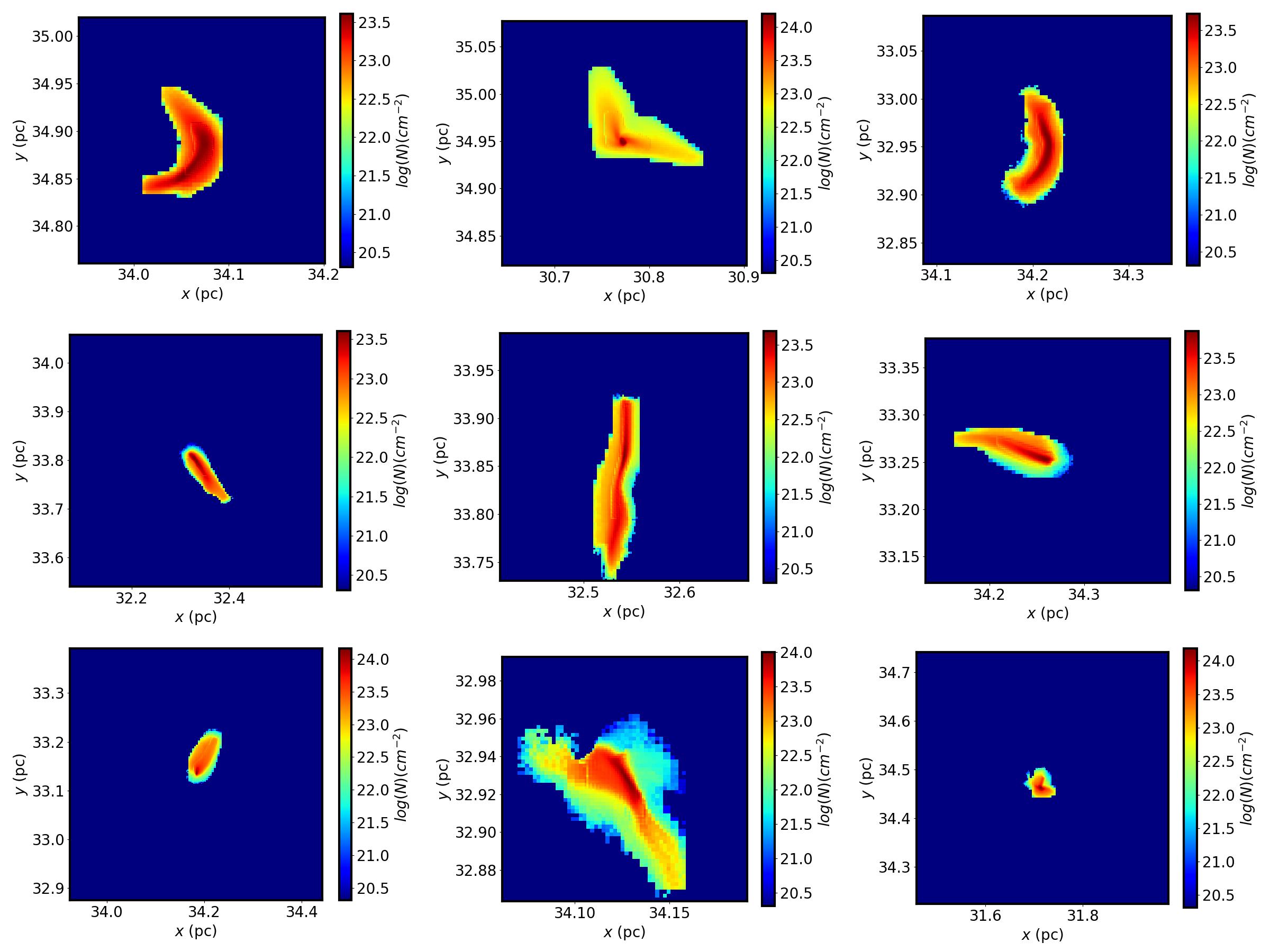}
   \caption{Column density of the cores shown in Fig. \ref{core_projections_mhd}, showing only the locations selected by the HOP algorithm.}
     \label{hierar_mhd}
\end{figure*}

In order to give a clearer idea of the core definitions, here we illustrate the core locations as selected by our implementation of the HOP algorithm. Figs. \ref{hierar_hydro} and \ref{hierar_mhd} show core projections on the xy plane, as in Figs. \ref{core_projections_hydro} and \ref{core_projections_mhd} of the text, but without the core environment.


\end{document}